\let\accentvec\vec  
\documentclass[number,sort&compress]{elsarticlesilent}
\let\vec\accentvec 

\usepackage{xspace}
\usepackage{amssymb}
\usepackage{graphicx} 	
\usepackage{enumerate}
\usepackage{url}
\usepackage{mathtools} 
\usepackage{amsthm}
\usepackage{multirow}
\usepackage{booktabs} 

\usepackage[shortlabels]{enumitem}
\setlist{noitemsep}

\usepackage{tikz}
\usetikzlibrary{shapes}

\graphicspath{{./images/}}

\usepackage[pdftex,                
            bookmarks=true,        
            bookmarksnumbered=true,
            bookmarksopen=false,
            hypertexnames=false,
]{hyperref}
\makeatletter
\providecommand*{\toclevel@title}{0}
\providecommand*{\toclevel@author}{0}
\makeatother

\makeatletter
\providecommand*{\toclevel@title}{0}
\providecommand*{\toclevel@author}{0}
\makeatother

\newcommand{\problemdef}[3]
{
\begin{quote}
\textsc{#1}\\
\textbf{Input:} #2\\
\textbf{Question:} #3
\end{quote}
}

\newcommand{\parproblemdef}[4]
{
\begin{quote}
\textsc{#1}\\
\textbf{Input:} #2\\
\textbf{Parameter:} #3\\
\textbf{Question:} #4
\end{quote}
}

\newcommand{\myqed}{}
\newcommand{\join}[0]{\textsc{join}\xspace}
\newcommand{\union}[0]{\textsc{union}\xspace}
\newcommand{\yes}[0]{\textsc{yes}\xspace}
\newcommand{\no}[0]{\textsc{no}\xspace}
\newcommand{\true}[0]{\textsc{true}\xspace}
\newcommand{\false}[0]{\textsc{false}\xspace}

\newcommand{\CG}[0]{\ensuremath{\mathop{\mathrm{\textsc{CG}}}}\xspace}
\newcommand{\joinheight}[0]{\ensuremath{\mathop{\mathrm{\textsc{JoinH}}}}\xspace}
\newcommand{\leftc}[0]{\ensuremath{\mathop{\mathrm{\textsc{left}}}}\xspace}
\newcommand{\rightc}[0]{\ensuremath{\mathop{\mathrm{\textsc{right}}}}\xspace}

\newcommand{\NPc}[0]{\ensuremath{\mathop{\mathrm{NP}}}-complete\xspace}
\newcommand{\coNPc}[0]{\ensuremath{\mathop{\mathrm{coNP}}}-complete\xspace}
\newcommand{\collapse}[0]{PH~$= \Sigma _3 ^p$\xspace}
\newcommand{\containment}[0]{NP~$\subseteq$~coNP$/$poly\xspace}
\newcommand{\ncontainment}[0]{NP~$\not \subseteq$~coNP$/$poly\xspace}
\newcommand{\f}[0]{\ensuremath{\mathcal{F}}\xspace}
\newcommand{\T}[0]{\ensuremath{\mathcal{T}}\xspace}
\newcommand{\fkv}[0]{\ensuremath{\mathcal{F}+k\mathrm{v}}\xspace}
\newcommand{\fkqv}[0]{\ensuremath{\mathcal{F}+(k+q)\mathrm{v}}\xspace}
\newcommand{\fpluske}[0]{\ensuremath{\mathcal{F}+k\mathrm{e}}\xspace}
\newcommand{\fminuske}[0]{\ensuremath{\mathcal{F}-k\mathrm{e}}\xspace}

\newcommand{\Gbar}[0]{\ensuremath{\overline{G}}\xspace}
\newcommand{\linearforest}[0]{\ensuremath{\mathrm{\textsc{Linear\-Forest}}}\xspace}

\newcommand{\cochordal}[0]{\ensuremath{\mathrm{\textsc{Cochordal}}}\xspace}
\newcommand{\ccochordal}[0]{\ensuremath{\mathrm{\textsc{$\bigcup$Cochordal}}}\xspace}
\newcommand{\ccochordalkv}[0]{\ensuremath{\mathrm{\textsc{$\bigcup$Cochordal}}+k\mathrm{v}}\xspace}

\newcommand{\chordalpluske}[0]{\ensuremath{\mathrm{\textsc{Chordal}}+k\mathrm{e}}\xspace}
\newcommand{\chordalminuske}[0]{\ensuremath{\mathrm{\textsc{Chordal}}-k\mathrm{e}}\xspace}
\newcommand{\cograph}[0]{\ensuremath{\mathrm{\textsc{Cograph}}}\xspace}
\newcommand{\bipartitekv}[0]{\ensuremath{\mathrm{\textsc{Bipartite}}+k\mathrm{v}}\xspace}
\newcommand{\bipartitepluske}[0]{\ensuremath{\mathrm{\textsc{Bipartite}}+k\mathrm{e}}\xspace}
\newcommand{\csplit}[0]{\ensuremath{\mathrm{\textsc{$\bigcup$Split}}}\xspace}
\newcommand{\csplitkv}[0]{\ensuremath{\mathrm{\textsc{$\bigcup$Split}}+k\mathrm{v}}\xspace}
\newcommand{\splitpluske}[0]{\ensuremath{\mathrm{\textsc{Split}}+k\mathrm{e}}\xspace}
\newcommand{\splitminuske}[0]{\ensuremath{\mathrm{\textsc{Split}}-k\mathrm{e}}\xspace}
\newcommand{\splitgraphs}[0]{\ensuremath{\mathrm{\textsc{Split}}}\xspace}
\newcommand{\splitkv}[0]{\ensuremath{\mathrm{\textsc{Split}}+k\mathrm{v}}\xspace}
\newcommand{\linearforestpmke}[0]{\ensuremath{\mathrm{\textsc{Linear\-Forest}}\pm k\mathrm{e}}\xspace}
\newcommand{\linearforestkv}[0]{\ensuremath{\mathrm{\textsc{Linear\-Forest}}+k\mathrm{v}}\xspace}

\newcommand{\pathgraphs}[0]{\ensuremath{\mathrm{\textsc{Path}}}\xspace}
\newcommand{\pathkv}[0]{\ensuremath{\mathrm{\textsc{Path}}+k\mathrm{v}}\xspace}
\newcommand{\forestkv}[0]{\ensuremath{\mathrm{\textsc{Forest}}+k\mathrm{v}}\xspace}
\newcommand{\intervalkv}[0]{\ensuremath{\mathrm{\textsc{Interval}}+k\mathrm{v}}\xspace}
\newcommand{\chordalkv}[0]{\ensuremath{\mathrm{\textsc{Chordal}}+k\mathrm{v}}\xspace}
\newcommand{\dominated}[0]{\ensuremath{\mathrm{\textsc{Dominated}}}\xspace}
\newcommand{\dominatedkv}[0]{\ensuremath{\mathrm{\textsc{Dominated}}+k\mathrm{v}}\xspace}
\newcommand{\simpledominated}[0]{\ensuremath{\mathrm{\textsc{Windmill}}}\xspace}
\newcommand{\simpledominatedkv}[0]{\ensuremath{\mathrm{\textsc{Windmill}}+k\mathrm{v}}\xspace}

\newcommand{\cographkv}[0]{\ensuremath{\mathrm{\textsc{Cograph}}+k\mathrm{v}}\xspace}
\newcommand{\emptykv}[0]{\ensuremath{\mathrm{\textsc{Independent}}+k\mathrm{v}}\xspace}
\newcommand{\independent}[0]{\ensuremath{\mathrm{\textsc{Independent}}}\xspace}
\newcommand{\completekv}[0]{\ensuremath{\mathrm{\textsc{Complete}}+k\mathrm{v}}\xspace}

\newcommand{\chromaticnumber}[0]{\ensuremath{\mathrm{\textsc{Chromatic Number}}}\xspace}
\newcommand{\dotcup}{\mathop{\dot\cup}}

\newcommand{\Oh}[0]{\ensuremath{\mathcal{O}}\xspace}

\newlength{\baseImageHeight}
\newcommand{\domset}[0]{\textsc{Dominating Set}\xspace}
\newcommand{\precoloringextension}[0]{\textsc{Precoloring Extension}\xspace}
\newcommand{\listcoloring}[0]{\textsc{List Coloring}\xspace}
\newcommand{\threelistcoloring}[0]{\textsc{$3$-List Coloring}\xspace}
\newcommand{\qlistcoloring}[0]{\textsc{\texorpdfstring{$q$}{q}-List Coloring}\xspace}
\newcommand{\qminlistcoloring}[0]{\textsc{\texorpdfstring{$(q-2)$}{q-2}-List Coloring}\xspace}
\newcommand{\chromaticnr}[0]{\textsc{Chromatic Number}\xspace}

\newcommand{\CNFSAT}[0]{\textsc{cnf-sat}\xspace}
\newcommand{\qsat}[0]{\textsc{$q$-cnf-sat}\xspace}
\newcommand{\tsat}[0]{\textsc{$t$-cnf-sat}\xspace}
\newcommand{\qnaesat}[0]{\textsc{$q$-nae-sat}\xspace}
\newcommand{\qcoloring}[0]{\textsc{$q$-Coloring}\xspace}
\newcommand{\qppnaesat}[0]{\textsc{$(q+1)$-nae-sat}\xspace}
\newcommand{\qppcoloring}[0]{\textsc{$(q+1)$-Coloring}\xspace}

\newcommand{\threecoloring}[0]{\textsc{$3$-Colo\-ring}\xspace}
\newcommand{\fourcoloring}[0]{\textsc{$4$-Coloring}\xspace}

\newtheorem{lemma}{Lemma}
\newtheorem{theorem}{Theorem}
\newtheorem*{cleaningRule}{Cleaning Rule}
\newtheorem{corollary}{Corollary}
\newtheorem*{claim}{Claim}
\newtheorem{proposition}{Proposition}

\theoremstyle{definition}
\newtheorem{definition}{Definition}

\newcommand{\sectref}[1]{Section~\ref{#1}}

\newcommand{\defref}[1]{Definition~\ref{#1}}
\newcommand{\lemmaref}[1]{Lemma~\ref{#1}}

\newcommand{\thmref}[1]{Theorem~\ref{#1}}
\newcommand{\corollaryref}[1]{Corollary~\ref{#1}}

\newcommand{\imgref}[1]{Fig.~\ref{#1}}

\newcommand{\proposref}[1]{Proposition~\ref{#1}}

\newcommand{\tableref}[1]{Table~\ref{#1}}

\begin{document}

\begin{frontmatter}

\title{Data Reduction for Graph Coloring Problems\tnoteref{t1,t2}}

\tnotetext[t1]{This work was supported by the Netherlands Organization for Scientific Research (NWO), project ``KERNELS: Combinatorial Analysis of Data Reduction''.}
\tnotetext[t2]{A preliminary version of this work appeared in the proceedings of the 18th International Symposium on Fundamentals of Computation Theory (FCT 2011). In addition to complete proofs, this full version contains several new results as explained in the introduction.}

\author[uu]{Bart M.\ P.\ Jansen\corref{cor1}}
\ead{B.M.P.Jansen@uu.nl}

\author[tub]{Stefan Kratsch\fnref{fn1}}
\ead{Stefan.Kratsch@tu-berlin.de}

\fntext[fn1]{Research done while at Utrecht University.}


\cortext[cor1]{Corresponding author}

\address[uu]{Utrecht University, P.O.\ Box 80.089, 3508 TB Utrecht, The Netherlands}

\address[tub]{Technical University Berlin, Germany}

\hypersetup{bookmarksdepth=-1}

\begin{abstract}
This paper studies the kernelization complexity of graph coloring problems with respect to certain structural parameterizations of the input instances. We are interested in how well polynomial-time data reduction can provably shrink instances of coloring problems, in terms of the chosen parameter. It is well known that deciding~$3$-colorability is already \NPc, hence parameterizing by the requested number of colors is not fruitful. Instead, we pick up on a research thread initiated by Cai (DAM, 2003) who studied coloring problems parameterized by the modification distance of the input graph to a graph class on which coloring is polynomial-time solvable; for example parameterizing by the number~$k$ of vertex-deletions needed to make the graph chordal. We obtain various upper and lower bounds for kernels of such parameterizations of \qcoloring, complementing Cai's study of the time complexity with respect to these parameters. 
Our results show that the existence of polynomial kernels for \qcoloring parameterized by the vertex-deletion distance to a graph class \f is strongly related to the existence of a function~$f(q)$ which bounds the number of vertices which are needed to preserve the \no-answer to an instance of \qlistcoloring on \f.
\end{abstract}

\end{frontmatter}

\hypersetup{bookmarksdepth=2}

\section{Introduction}
Graph coloring is one of the most well-studied and well-known topics in graph algorithmics and discrete mathematics; it hardly needs an introduction. In this work we study the kernelization complexity of graph coloring problems, or in other words the existence of efficient and provably effective preprocessing procedures, using the framework of parameterized complexity~\cite{DowneyF99,GuoN07a} (consult \sectref{preliminaries} for full definitions). Parameterized complexity enables us to study qualitatively and quantitatively how different properties of a graph coloring instance contribute to its difficulty.

The choice of parameter is therefore very important. If we consider the vertex coloring problem and parameterize by the requested number of colors, then this problem is already \NPc for a constant value of~$3$ for the parameter~\cite[GT4]{GareyJ79}, resulting in intractability; we should consider different parameterizations to obtain meaningful questions. In his study of the parameterized complexity of vertex coloring problems, Leizhen Cai~\cite{Cai03a} introduced a convenient notation to talk about \emph{structural} parameterizations of graph problems. For a graph class~$\f$ let~$\fkv$ denote the graphs which can be built by adding at most~$k$ vertices to a graph in~$\f$; the neighborhoods of these new vertices can be arbitrary. Equivalently the class~$\fkv$ contains those graphs which contain a \emph{modulator}~$X \subseteq V(G)$ of size at most~$k$ such that~$G - X \in \f$. Hence~$\forestkv$ is exactly the class of graphs which have a feedback vertex set of size at most~$k$. Similarly one may define classes~$\fpluske$ and~$\fminuske$ where the structure is measured through the number of \emph{edges} which were added or removed from a member of~$\f$ to build the graph. Using this notation we can define a class of parameterized coloring problems with structural parameters.
\parproblemdef{$q$-Coloring on \fkv graphs}
{An undirected graph~$G$ and a modulator~$X \subseteq V(G)$ such that $G - X \in \f$.}
{The size~$k := |X|$ of the modulator.}
{Is $\chi(G) \leq q?$}
To decouple the existence of polynomial kernelizations from the difficulties of \emph{finding} a modulator, we assume that a modulator is given in the input. The \chromaticnr \textsc{on} \fkv graphs problem is defined similarly as \qcoloring, with the important exception that the value~$q$ is not fixed, but part of the input. For the purposes of kernelization, however, there is little left to explore for \chromaticnr: a superset of the authors showed~\cite[Theorem 14]{BodlaenderJK11} that \chromaticnr does not admit a polynomial kernel when parameterized by the vertex cover number, or in Cai's notation: \chromaticnr on \emptykv graphs does not admit a polynomial kernel, unless \containment and the polynomial hierarchy collapses to the third level~\cite{Yap83} (\collapse). The proof given in that paper shows that even a compound parameterization by the vertex cover number \emph{plus} the number of colors that is asked for, does not admit a polynomial kernel. Hence it seems that the size of the kernel must depend super-polynomially on the number of colors. In this work we therefore focus on \qcoloring and consider how the structural parameterizations influence the complexity of the problem when keeping the number of colors~$q$ \emph{fixed}.


\tikzset{param/.style={
rounded rectangle,
minimum width=0.8cm,
minimum height=0.3cm,
very thick,
draw=black!90!white!90, 
top color=white, 
bottom color=red!50!black!20, 
text width=1.7cm,
rounded corners=2pt,
text centered 
}}

\tikzset{legend/.style={
rectangle,
minimum width=0.8cm,
minimum height=0.2cm,
very thick,
draw=black!90!white!90, 
top color=white, 
bottom color=red!50!black!20, 
rounded corners=2pt,
rounded corners=2pt,
}}

\tikzset{paraNPC/.style={
very thick,
top color=white,
bottom color=black!50!white,
}}

\tikzset{FPT/.style={
thin,
top color=white,
bottom color=black!10!white,
}}

\tikzset{WhardInXP/.style={
very thick,
top color=white,
bottom color=black!20!white,
dashed
}}

\tikzset{InXP/.style={
very thick,
dotted,
top color=white,
bottom color=black!30!white,
}}

\tikzset{fptNoKernels/.style={
bottom color=black!30!red!60, top color=white
}}

\tikzset{fptWithKernels/.style={
top color=green!30, bottom color=white
}}

\begin{figure}[t]
	\small

	\centering
\begin{tikzpicture}[inner sep=0.1cm, x={(0cm,-1.3cm)}, y={(2.5cm, 0cm)}, >=stealth, thick]

\draw (0-0.5,-3) [rounded corners=8pt, fptNoKernels] rectangle (4-0.6, 1.5);

\draw (0-0.5 + 0.04, -0.5) [rounded corners=8pt, fptWithKernels] rectangle (2-0.25 - 0.12 + 1 - 0.2, 1.5 - 0.02);

\node at (-0.1,-3 + 0.9) {\qcoloring~$\in$~FPT};

\node at (-0.1, 1) [text width=2cm, text centered] {Polynomial kernels};

\node at ( 0,0) [param,FPT] (empty) {$\star$ \textsc{Indepen\-dent} + $k$v};
\node at ( 1,1) [param,InXP] (cograph) {$\star$ \textsc{Cograph} + $k$v \cite{KoblerR03,FominGLS10}};
\node at ( 1,0) [param,WhardInXP] (split) {$\star$ \textsc{$\bigcup$Split} + $k$v~\cite{Cai03a}};
\node at ( 2,-1) [param,WhardInXP] (interval) {\textsc{Interval} + $k$v~\cite{Marx06a}};
\node at ( 3,0) [param,WhardInXP] (chordal) {\textsc{Chordal} + $k$v~\cite{Marx06a}};

\node at ( 2,0.5) [param,WhardInXP] (cochordal) {\textsc{$\bigcup$Cochor\-dal}+$k$v $\star$};

\node at ( 2,-2) [param,FPT] (forest) {\textsc{Forest} + $k$v};
\node at ( 1,-1) [param,FPT] (linforest) {$\star$ \textsc{Linear\-Forest}+$k$v};
\node at ( 3.5+0.6,-1) [param,paraNPC] (bipartite) {\textsc{Bipartite} + $k$v \cite{Kratochvil93,Cai03a,BodlaenderJW94}};
\node at ( 3,-2.5) [param,FPT] (treewidth) {\textsc{Treewidth} \cite{BodlaenderK08}};
\node at ( 4+0.5,-2.5) [param,paraNPC] (chromaticnr) {\textsc{Chromatic Number}};
\node at ( 4+0.5,1) [param,paraNPC] (perfect) {\textsc{Perfect} + $k$v};

\draw [->] (empty) -- (cograph);
\draw [->] (empty) -- (split);
\draw [->] (empty) -- (linforest);
\draw [->] (cograph) -- (perfect);
\draw [->] (split) -- (chordal);
\draw [->] (split) -- (cochordal);
\draw [->] (cochordal) -- (perfect);
\draw [->] (linforest) -- (forest);
\draw [->] (linforest) -- (interval);
\draw [->] (interval) -- (chordal);
\draw [->] (chordal) -- (perfect);
\draw [->] (forest) -- (bipartite);
\draw [->] (bipartite) -- (perfect);
\draw [->] (bipartite) -- (chromaticnr);
\draw [->] (forest) -- (treewidth);
\draw [->] (treewidth) -- (chromaticnr);

\end{tikzpicture}

	\caption{The hierarchy of parameters used in this work. Arrows point from larger parameters to smaller parameters: an arc~$P \rightarrow P'$ signifies that every graph~$G$ satisfies~$P(G) + 2 \geq P'(G)$. For parameters with a star we obtain new parameterized complexity results in this work.
	The complexity of the \chromaticnumber problem for a given parameterization is expressed through the shading and border style of the parameters: the complexity status can be NP-complete for fixed~$k$
\protect\tikz{\protect \node at (0,0) [legend,paraNPC] {};},
or contained in XP but not known to be W$[1]$-hard
\protect\tikz{\protect \node at (0,0) [legend,InXP] {};},
or contained in XP and W$[1]$-hard 
\protect\tikz{\protect \node at (0,0) [legend,WhardInXP] {};},
or FPT but without polynomial kernel unless \containment
\protect\tikz{\protect \node at (0,0) [legend,FPT] {};} 
. The complexity of \qcoloring for a given parameterization is expressed through the containers: the status is either FPT with a polynomial kernel, FPT but no polynomial kernel unless \containment, or NP-complete for fixed~$k$.
	}
	\label{ParameterHierarchy}
\end{figure}
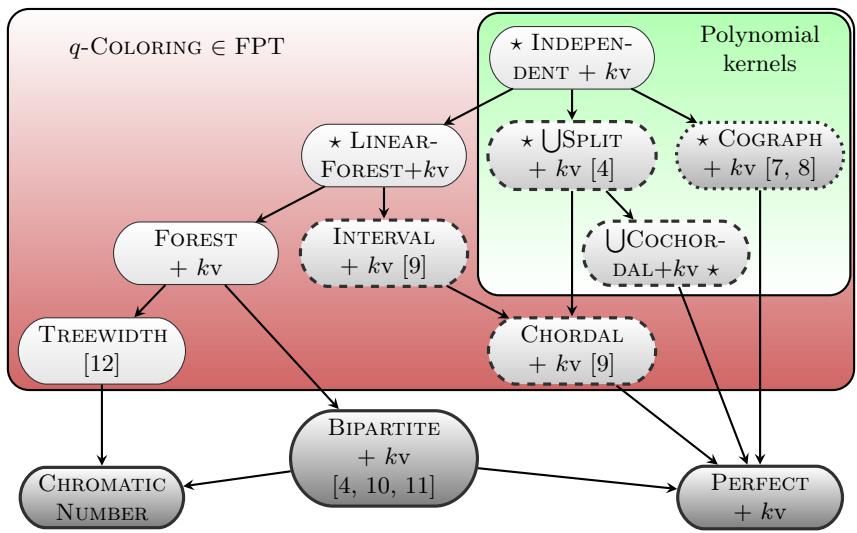

When studying coloring problems with these structural parameterizations, we feel that it is important to look at the relations between the parameters and consider the parameter space as a \emph{hierarchy} (\imgref{ParameterHierarchy}), rather than exploring a table of problems versus parameters one row at a time (cf.~\cite{FellowsLMMRS09}). It is known that there are several coloring problems such as \textsc{Precoloring Extension} and \textsc{Equitable Coloring} which are W$[1]$-hard when parameterized by treewidth, but fixed-parameter tractable parameterized by the vertex cover number~\cite{FialaGK11,FellowsFLRSST11}. These parameters also yield differences in the \emph{kernelization complexity} of \qcoloring. Our hierarchy includes these parameters, and several others which are sandwiched between them.

\textbf{Our results.}
In this paper we pinpoint the boundary for polynomial kernelizability of \qcoloring in the given hierarchy, by exhibiting upper- and lower bounds for kernel sizes. For all parameters in \imgref{ParameterHierarchy} for which \qcoloring is in FPT we either give a polynomial kernel, or prove that the existence of such a kernel would imply \containment and is therefore unlikely.

\emph{Upper bounds in the hierarchy.} We derive a general theorem which associates the existence of polynomial kernels for \qcoloring on \fkv graphs to properties of the \qlistcoloring problem on graphs in \f: if the non-list-colorability of a graph in \f is ``local'' in the sense that for any \no-instance there is a small subinstance on~$g(q)$ vertices to which the answer is \no, then \qcoloring on \fkv graphs admits a polynomial kernel with $\Oh(k^{q \cdot g(q)})$ vertices for every fixed~$q$. This positive news even extends to the \qlistcoloring problem with structural parameterizations. We apply the general theorem to give polynomial kernels for \qcoloring on the parameterized graph families \cographkv and \ccochordalkv. The latter is a strict generalization of a result of the preliminary version, where we obtained kernels on \csplitkv graphs: observe that $\csplit$ graphs, the unions of split graphs, are contained in the class $\ccochordal$, the unions of cochordal graphs, since the complement of every split graph is chordal.

\emph{Lower bounds in the hierarchy.} In the seminal paper on kernelization lower-bounds, Bodlaender et al.~\cite[Theorem 2]{BodlaenderDFH09} prove that \threecoloring parameterized by treewidth does not admit a polynomial kernel unless all \coNPc problems have distillation algorithms. We strengthen their result by showing that unless \containment (an even less likely condition), the problem does not even admit a polynomial kernel parameterized by vertex-deletion distance to a single path: \threecoloring on \pathkv graphs does not admit a polynomial kernel. Under the same assumption, this immediately excludes polynomial kernels on e.g.\ \forestkv or \intervalkv graphs, since the latter are \emph{smaller} parameters. 

Besides investigating the \emph{existence} of polynomial kernels, we also study the \emph{degree} of the polynomial in the kernels that we obtain for \qcoloring on \fkv graphs. Our general scheme yields kernels with~$\Oh(k^q)$ vertices on \emptykv graphs, and a small insight allows us to encode these instances in~$\Oh(k^q)$ bits. Using a connection to \textsc{Not-All-Equal $q$-Satisfi\-ability} (\qnaesat) we prove that for every~$q \geq 4$ the \qcoloring problem parameterized by vertex cover does not admit a kernel which can be encoded in~$\Oh(k^{q - 1 - \varepsilon})$ bits for any~$\varepsilon > 0$ unless \containment. 

As a new result in this full version we show that the properties of \no-instances to \qlistcoloring on graphs from \f cannot just be used to obtain kernel size \emph{upper bounds} for \qcoloring on \fkv graphs, but also \emph{lower bounds}. We prove that if there is a \no-instance of \qminlistcoloring on~$t$ vertices which is irreducible in the sense that any vertex-deletion turns it into a \yes-instance, then kernels for \qcoloring on \fkv graphs cannot have bitsize~$\Oh(k^{t - \varepsilon})$ for any~$\varepsilon > 0$ unless \containment. This theorem explains why the number of vertices in our general kernelization scheme must depend on the function~$g(q)$ in its exponent.

\emph{Domination-related parameters.} It turns out that the difficulty of a \threecoloring instance is intimately related to the domination-properties of the graph. We show the surprising (but not difficult) result that \threecoloring on a general graph~$G$ can be solved in~$\Oh^*(3^k)$ time when a dominating set~$X$ of size~$k$ is given along with the input. In contrast we show that \threecoloring parameterized by the size of a dominating set does not admit a polynomial kernel unless \containment. To obtain polynomial kernels by exploiting the domination structure of the graph, we must consider another parameter. Let \dominated be the graphs where each connected component has a dominating vertex. We show that \threecoloring on \dominatedkv graphs admits a polynomial kernel. This cannot be extended to arbitrary~$q$, since \fourcoloring is \NPc on \dominated graphs.

\textbf{Related work.} Structural parameterizations of graph coloring problems were first studied by Cai~\cite{Cai03a} who showed, amongst others, that \chromaticnr is in FPT on \chordalminuske, \splitpluske, and \splitminuske graphs, but W$[1]$-hard on \splitkv graphs, and that \threecoloring on \bipartitekv and \bipartitepluske graphs is \NPc for constant values of~$k$. Marx~\cite{Marx06a} continued this line of research and showed e.g.\ that \chromaticnr on \intervalkv and \chordalkv graphs is in XP but W$[1]$-hard, and that \chromaticnr on \chordalpluske graphs is in FPT. A summary of their results relevant to this work can be found in \imgref{ParameterHierarchy}. Chor, Fellows, and Juedes~\cite{ChorFJ04} considered the problem of coloring a graph on~$n$ vertices with~$n - k$ colors and obtained an FPT algorithm. They also showed that the problem of covering the vertices of a graph with $n - k$ cliques has a kernel with~$3k$ vertices. Using the fact that \chromaticnr on \completekv graphs is equivalent to partitioning the complement graph into cliques parameterized by vertex cover, and that a graph~$G$ with vertex cover of size~$k$ needs at least~$n - k$ cliques to cover it, their result implies that \chromaticnr on \completekv graphs has a linear-vertex kernel. Finally we observe that the \qcoloring problem on \intervalkv and \chordalkv graphs is in FPT, using the fact that either the interval- or chordal graph~$G - X$ contains a clique of size~$q + 1$ and the answer is \no, or the treewidth of~$G$ is at most~$k + q$. An algorithm to solve \chromaticnr on \ccochordalkv graphs in XP-time can be obtained using the ideas of \lemmaref{lemma:coChordalSmallCertificates}.

\textbf{Organization.} The outline of the paper follows the description of our results. We give preliminaries on parameterized complexity and graph theory in \sectref{preliminaries}. Positive results related to the hierarchy are established in \sectref{section:positiveResults}, by first deriving general theorems in \sectref{section:kernelUpperBounds} and then establishing upper bounds on the sizes of \no-certificates for \qlistcoloring in \sectref{section:certificateUpperBounds} to use in these theorems. In \sectref{section:negativeResults} we focus on negative results in the parameter hierarchy, giving lower bounds for important parameterizations in \sectref{section:kernelLowerBounds}, finishing with a general negative theorem which is applied using the lower bound constructions for sizes of \no-certificates which are developed in \sectref{section:certificateLowerBounds}. We treat parameterizations related to the domination-structure of the graph in \sectref{section:dominationRelated}.

\section{Preliminaries} \label{preliminaries}
\textbf{Parameterized complexity and kernels.}
A parameterized problem~$Q$ is a subset of~$\Sigma^* \times \mathbb{N}$, the second component being the \emph{parameter} which expresses some structural measure of the input. A parameterized problem is (strongly uniform) \emph{fixed-parameter tractable} if there exists an algorithm to decide whether $(x,k) \in Q$ in time~$f(k)|x|^{\Oh(1)}$ where~$f$ is a computable function.

A \emph{kernelization algorithm} (or \emph{kernel}) for a parameterized problem~$Q$ is a polynomial-time algorithm which transforms an instance~$(x,k)$ into an equivalent instance~$(x', k')$ such that~$|x'|, k' \leq f(k)$ for some computable function~$f$, which is the \emph{size} of the kernel. If~$f \in k^{\Oh(1)}$ is a polynomial then this is a \emph{polynomial kernel}. Intuitively a polynomial kernel for a parameterized problem is a polynomial-time preprocessing procedure which reduces the size of an instance to something which \emph{only} depends (polynomially) on the parameter~$k$, and does not depend at all on the input size~$|x|$.

\begin{definition}[cf. \cite{Bodlaender09}] \label{def:polyParameterTransform}
Let~$P$ and~$Q$ be parameterized problems. We say that~$P$ is \emph{polynomial-parameter reducible} to~$Q$, written~$P \leq_{ptp} Q$, if there exists a polynomial time computable function~$f \colon \Sigma^* \times \mathbb{N} \to \Sigma^* \times \mathbb{N}$, and a polynomial~$p \colon \mathbb{N} \to \mathbb{N}$ such that for all~$(x, k) \in \Sigma^* \times \mathbb{N}$ it holds that (a) $(x,k) \in P$ if and only if $(x', k') = f(x,k) \in Q$ and (b) $k' \leq p(k)$. The function~$f$ is called a \emph{polynomial-parameter transformation}. In the special case that~$p$ is a linear function we have a \emph{linear-parameter transformation}.
\end{definition}
Under a mild technical assumption which is satisfied by all applications in this work, a polynomial-parameter transformation from~$P$ to~$Q$, together with a polynomial kernel for~$Q$, yields a polynomial kernel for~$P$~\cite{Bodlaender09}. Using contraposition this can be exploited to give kernel lower-bounds for~$Q$ based on lower-bounds for~$P$. We also use a compression lower-bound for \qsat due to Dell and van Melkebeek, who built on a theorem by Fortnow and Santhanam~\cite{FortnowS11}. Although the result~\cite[Theorem 1]{DellM10} was originally phrased in terms of an oracle communication protocol, we only need the following weaker statement.
\begin{proposition} [\cite{DellM10}] \label{dellMelkebeekCompressionBound}
If \ncontainment then for any~$\varepsilon > 0$ and~$q \geq 3$ there is no polynomial-time algorithm which transforms an instance~$x$ of \qsat on~$n$ variables into an equivalent instance~$x'$ of a decidable problem with bitsize~$|x'| \in \Oh(n^{q - \varepsilon})$.
\end{proposition}
We also use the following corollary to this statement.
\begin{proposition}[\cite{DellM10}] \label{proposition:cnfSatNoPoly}
\CNFSAT parameterized by the number of variables does not admit a polynomial kernel unless \containment.
\end{proposition}

\textbf{Graphs.}
All graphs we consider are finite, undirected and simple. We use $V(G)$ and~$E(G)$ to denote the vertex- and edge set of a graph~$G$. For~$X \subseteq V(G)$ the subgraph induced by~$X$ is denoted by~$G[X]$. The terms $P_n$, $C_n$, and~$K_n$ denote the path, cycle, and complete graphs on~$n$ vertices, respectively. For natural numbers~$q$ we define~$[q] := \{1, 2, \ldots, q\}$. A proper $q$-coloring of a graph~$G$ is a function~$f \colon V(G) \to [q]$ such that adjacent vertices receive different colors. The \emph{chromatic number}~$\chi(G)$ of a graph is the smallest number~$q$ for which it has a proper $q$-coloring. For a vertex set~$X \subseteq V(G)$ we denote by~$G - X$ the graph obtained from~$G$ by deleting all vertices of~$X$ and their incident edges. An odd cycle is a simple cycle on an odd number of at least three vertices. An odd wheel is the graph which is obtained from an odd cycle by adding a new vertex which is adjacent to all other vertices. A \emph{diamond} is the graph obtained from~$K_4$ by deleting a single edge; every edge yields the same result. A vertex~$v$ in a graph~$G$ is \emph{simplicial} if~$N_G(v)$ is a clique. If~$v$ is contained in the connected component~$C$, then~$v$ is a \emph{dominating vertex} for~$C$ if~$N_G[v] = V(C)$. The \emph{edge-complement} \Gbar of a graph~$G$ has the same vertex set as~$G$, with an edge between two vertices if and only if they are not adjacent in~$G$.

The \emph{join} of two graphs~$G_1$ and~$G_2$ is the graph~$G_1 \otimes G_2$ on the vertex set~$V(G_1) \dotcup \linebreak[1] V(G_2)$ and edge set~$E(G_1) \dotcup E(G_2) \dotcup \{ \{x,y\} \mid x \in V(G_1) \wedge y \in V(G_2) \}$. The \emph{union} of~$G_1$ and~$G_2$ is the graph~$G_1 \cup G_2$ with vertices~$V(G_1) \dotcup V(G_2)$ and edges~$E(G_1) \dotcup E(G_2)$.

A \emph{graph class} is a (possibly infinite) set of graphs. A graph class is \emph{hereditary} if the class is closed under taking vertex-induced subgraphs. A graph~$G$ with~$E(G) = \emptyset$ is called an \emph{independent} graph. A graph is a \emph{split graph} if there is a partition of~$V(G)$ into sets~$X, Y$ such that~$X$ is a clique and~$Y$ is an independent set. We define the class \csplit containing all disjoint-unions of split graphs. A graph is a \emph{cograph} if it does not contain~$P_4$ as an induced subgraph. A graph is \emph{chordal} if every cycle of length at least four has a chord, i.e., an edge between two vertices that are not successive on the cycle. A \emph{cochordal} graph is the edge-complement of a chordal graph, and \ccochordal is the class of graphs in which each connected component is cochordal. A \emph{linear forest} is a disjoint union of paths. The book by Brandst\"adt, Le, and Spinrad~\cite{BrandstadtLS99} contains more information about the graph classes used in this work.

For a finite set~$X$ and non-negative integer~$i$ we write~$\binom{X}{i}$ for the collection of all size-$i$ subsets of~$X$, ~$\binom{X}{\leq i}$ for all size \emph{at most}~$i$ subsets of~$X$, and~$X^i$ for the Cartesian product of~$i$ copies of~$X$.

\begin{definition} \label{def:cotree}
A \emph{cotree}~$\T$ is a rooted proper binary tree whose internal vertices are labeled as \join or \union nodes. For~$v \in V(\T)$ the graph~$\CG(\T, v)$ represented by the subtree of~$\T$ rooted at~$v$ is defined as follows:
\begin{equation} \nonumber
\CG(\T, v) := \begin{cases}
\mbox{Graph~$G := (\{v\}, \emptyset)$} & \mbox{If~$v$ is a leaf.} \\
\CG(\T, \leftc(v)) \cup \CG(\T, \rightc(v)) & \mbox{If~$v$ is a \union node.} \\
\CG(\T, \leftc(v)) \otimes \CG(\T, \rightc(v)) & \mbox{If~$v$ is a \join node.}
\end{cases}
\end{equation}
where~$\leftc(v)$ and~$\rightc(v)$ are the left- and right child of node~$v$ in the tree~$\T$, respectively. We define~$\CG(\T)$ as~$\CG(\T, r)$, where~$r$ is the root of~$\T$, and we say that~$\T$ is a \emph{cotree-representation} of the graph~$\CG(\T)$. The \emph{\join-height} $\joinheight(\T)$ of a cotree~$T$ is the maximum number of \join nodes on any path from the root to a leaf.
\end{definition}
It is well-known that a graph is a cograph if and only if it has a cotree-representation~\cite{BrandstadtLS99}. Let~$\omega(G)$ denote the size of the largest clique in~$G$. Using the fact that~$\omega(G_1 \otimes G_2) = \omega(G_1) + \omega(G_2)$, it is not hard to verify the following proposition.
\begin{proposition} \label{joinHeight}
If~$\T$ is a cotree representation of~$G$ then~$\omega(G) \geq \joinheight(\T) + 1$. 
\end{proposition}
The following facts will be useful in the remainder of the paper.
\begin{proposition}[{\cite[Chapter 1.2]{BrandstadtLS99}}] \label{proposition:chordalHasSimplicial}
Every chordal graph has a simplicial vertex.
\end{proposition}
\begin{proposition} \label{twoColoringAPath}
In any proper $2$-coloring of a graph~$P_{2n}$, the first and last vertex on the path must receive a different color.
\end{proposition}

\section{Positive results in the hierarchy} \label{section:positiveResults}
\subsection{Kernelization upper bounds} \label{section:kernelUpperBounds}
In this section we present several positive results (i.e., polynomial kernels) for parameterizations in the considered hierarchy. We begin by giving a general theorem which proves the existence of polynomial kernels for \qcoloring on \fkv graphs under certain conditions on the class \f. In \sectref{section:certificateUpperBounds} we investigate which classes \f satisfy these conditions. We introduce some terminology related to the \qlistcoloring problem to state the general theorem and its proof precisely.
\problemdef{\qlistcoloring}
{An undirected graph~$G$ and for each vertex~$v \in V(G)$ a list~$L(v) \subseteq [q]$ of allowed colors.}
{Is there a proper $q$-coloring $f \colon V(G) \to [q]$ such that $f(v) \in L(v)$ for each~$v \in V(G)$?}
Observe that in our definition of \qlistcoloring, the \emph{total} color universe has size~$q$ and the lists can have any size. The same problem name is sometimes used for the variant in which there can be arbitrarily many colors while each list has size~$q$, but we do not consider this variant in this work.

Let us define an instance~$(G', L')$ of \qlistcoloring as a \emph{subinstance} of~$(G,L)$ if~$G'$ is an induced subgraph of~$G$ and~$L'(v) = L(v)$ for all~$v \in V(G')$. If~$(G',L')$ is a \no-instance we say it is a \no-subinstance. The main condition on \f which is needed to ensure the existence of polynomial kernels for \qcoloring is captured by the following definition.
\begin{definition} \label{def:smallNoCertificates}
Let~$g \colon \mathbb{N} \to \mathbb{N}$ be a function. Graph class \f has $g(q)$-size \no-certificates for \qlistcoloring if for all \no-instances~$(G,L)$ of \qlistcoloring with~$G \in \f$ there is a \no-subinstance~$(G',L')$ on at most~$g(q)$ vertices.
\end{definition}
We shall see later that such a bounding function~$g(q)$ can be found for the graph classes \cograph, \csplit, and \ccochordal, whereas no such bound exists for the class of all paths. The existence of small \no-certificates for \qlistcoloring on \f turns out to be intimately linked to the existence of polynomial kernels for \qcoloring on \fkv graphs, as is shown in the following theorem.
\begin{theorem} \label{generalKernel}
Let \f be a hereditary class of graphs with~$g(q)$-size \no-certificates for \qlistcoloring. The \qcoloring problem on \fkv graphs admits a polynomial kernel with $\Oh(k^{q \cdot g(q)})$ vertices for every fixed~$q$.
\end{theorem}
\begin{proof}
Consider an instance~$(G, X)$ of \qcoloring on a graph class \fkv which satisfies the stated requirements. We give an outline of the reduction algorithm.
\begin{enumerate}
	\item 
	\begin{description}
	\item[For each] undirected graph~$H$ on~$t \leq g(q)$ vertices~$\{h_1, \ldots, h_t\}$, do:
	\begin{description}
		\item[For each] tuple~$(S_1, \ldots, S_t) \in \binom{X}{\leq q}^t$, do:
		\begin{description}
			\item[If] there is an induced subgraph of~$G - X$ on vertices~$\{v_1, \ldots, v_t\}$ which is isomorphic to~$H$ by the mapping~$h_i \mapsto v_i$, and~$S_i \subseteq N_G(v_i)$ for~$i \in [t]$, then mark the vertices~$\{v_1, \ldots, v_t\}$ as \emph{important} for \emph{one} such subgraph, which can be chosen arbitrarily.
		\end{description}
	\end{description}
\end{description}
	\item Let~$Y$ contain all vertices of~$G - X$ which were marked as important, and output the instance~$(G', X)$ with~$G' := G[X \cup  Y]$.
\end{enumerate}
Let us verify that this procedure can be executed in polynomial time for fixed~$q$, and leads to a reduced instance of the correct size. The number of undirected graphs on~$g(q)$ vertices is constant for fixed~$q$. The number of considered tuples is bounded by~$\Oh \left((q|X|^q)^{g(q)} \right)$, and for each graph~$H$, for each tuple, we mark at most~$g(q)$ vertices which is a constant. These observations imply that the algorithm outputs an instance of the appropriate size, and that it can be made to run in polynomial time for fixed~$q$ because we can just try all possible isomorphisms by brute-force. Since~$G' - X$ is an induced subgraph of~$G - X$, it follows that~$G' - X \in \f$ because \f is hereditary. It remains to prove that the two instances are equivalent: $\chi(G) \leq q \Leftrightarrow \chi(G') \leq q$. The forward direction of this equivalence is trivial, since~$G'$ is a subgraph of~$G$. We now prove the reverse direction.

Assume that~$\chi(G') \leq q$ and let~$f' \colon V(G') \to [q]$ be a proper $q$-coloring of~$G'$. Obtain a partial $q$-coloring~$f$ of~$G$ by copying the coloring of~$f'$ on the vertices of~$X$. Since~$G'[X] = G[X]$ the function~$f$ is a proper partial $q$-coloring of~$G$, which assigns all vertices of~$X$ a color. We will prove that~$f$ can be extended to a proper $q$-coloring of~$G$, using an argument about list-coloring. Consider the graph~$H := G - X$ which contains exactly the vertices of~$G$ which are not yet colored by~$f$. For each vertex~$v \in V(H)$ define a list of allowed colors as~$L(v) := [q] \setminus \{ f(u) \mid u \in N_G(v) \}$, i.e.\ for every vertex we allow the colors which are not yet used on a colored neighbor in~$G$. From this construction it is easy to see that any proper $q$-list-coloring of the instance~$(H, L)$ gives a valid way to augment~$f$ to a proper $q$-coloring of all vertices of~$G$. Hence it remains to prove that~$(H,L)$ is a \yes-instance of \qlistcoloring. 

Assume for a contradiction that~$(H,L)$ is a \no-instance. Since the problem definition ensures that $H = G - X \in \f$ the assumptions on~$\f$ imply there is a \no-subinstance~$(H', L')$ on~$t \leq g(q)$ vertices.

Let the vertices of~$H'$ be~$h_1, \ldots, h_t$. Since~$(H',L')$ is a subinstance of~$(H,L)$ we know by construction of the latter that for every vertex~$h_i$ with~$i \in [t]$ and for every color~$j \in [q] \setminus L'(h_i)$ there is a vertex of~$N_G(h_i)$ which is colored with~$j$. Now choose sets~$S_1, \ldots, S_t$ such that for every~$j \in [q] \setminus L'(h_i)$ set~$S_i$ contains exactly one neighbor~$v \in N_G(h_i)$ with~$f(v) = j$, which is possible by the previous observation. Since~$f$ only colors vertices from~$X$ we have~$S_i \subseteq X$ for all~$i \in [t]$.

Because~$H'$ is an induced subgraph on at most~$g(q)$ vertices of~$H = G - X$, we must have considered graph~$H'$ during the outer loop of the reduction algorithm. Since each~$S_i$ contains at most~$q$ vertices from~$X$, we must have considered the tuple~$(S_1, \ldots, S_t)$ during the inner loop of the reduction algorithm, and because the existence of~$H'$ shows that there is at least one induced subgraph of~$G-X$ which satisfies the if-condition, we must have marked some vertices~$\{v_1, \ldots, v_t\}$ of an induced subgraph~$H^*$ of~$G-X$ isomorphic to~$H'$ by an isomorphism~$v_i \mapsto h_i$ as \emph{important}, and hence the induced subgraph~$H^*$ is contained in the graph~$G'$. Recall that~$f'$ is a proper $q$-coloring of~$G'$, and that~$f$ and~$f'$ assign the same colors to vertices of~$X$. By construction this shows that for each vertex~$h_i$ with~$i \in [t]$ of the presumed \no-subinstance $(H',L')$ of \qlistcoloring, for each color~$j \in [q] \setminus L'(h_i)$ which is not on the list of~$h_i$, there is a neighbor of the corresponding vertex~$v_i$ (i.e.\ a vertex in~$N_{G'}(v_i)$) which is colored~$j$. Using the fact that~$H^*$ is isomorphic to~$H'$ we obtain a valid $q$-list-coloring of~$H'$ by using the colors assigned to~$H^*$ by~$f'$. But this shows that~$(H',L')$ is in fact a \yes-instance of \qlistcoloring, which contradicts our initial assumption. This proves that the instance~$(H,L)$ of \qlistcoloring that we created must be a \yes-instance, and by our earlier observations this implies that~$\chi(G) \leq q$, which concludes the proof of the equivalence of the input- and output instance.

Hence we have shown that for each fixed~$q$ there is a polynomial-time algorithm which transforms an input of \qcoloring on \fkv graphs into an equivalent instance of size bounded polynomially in~$k = |X|$, which concludes the proof.
\myqed
\end{proof}

As an example application of \thmref{generalKernel}, observe that a \no-instance of \qlistcoloring on an independent graph has an induced \no-subinstance on a single vertex: as there are no edges, a \no-instance must have a vertex with an empty list. Since graphs without edges are hereditary, the proof of \thmref{generalKernel} gives the following corollary.

\begin{corollary} \label{qcoloringByVertexCover}
\qcoloring on \emptykv graphs (i.e.\ parameterized by vertex cover) admits a polynomial kernel with~$\Oh(k^q)$ vertices for every fixed integer~$q$.
\end{corollary}


While most kernels with~$\Oh(k^q)$ vertices require~$\Omega(k^{q + 1})$ bits to represent (as the number of edges can be quadratic in the number of vertices), we can prove that a kernel for \qcoloring on \emptykv graphs exists that can be encoded in~$\Oh(k^q)$ bits.

\begin{lemma} \label{smallVertexCoverKernel}
For every fixed~$q \geq 3$, \qcoloring on \emptykv graphs (i.e.\ parameterized by vertex cover) admits a kernel with~$k + k^q$ vertices that can be encoded in~$\Oh(k^q)$ bits.
\end{lemma}
\begin{proof}
Let~$(G, X)$ be an instance of \qcoloring on \emptykv graphs, which implies that~$X$ of size~$k$ is a vertex cover of~$G$. We create an equivalent instance~$(G', X)$ as follows. Start by setting~$G' := G[X]$. For every~$S \in \binom{X}{q}$, if~$\bigcap _{v \in S} N_G(v) \setminus X \neq \emptyset$, add a new vertex~$v_S$ to~$G'$ with~$N_{G'}(v_S) := S$. Using argumentation similar to that of \thmref{generalKernel} it can be proven that~$(G,X)$ is equivalent to the instance~$(G', X)$: from a proper~$q$-coloring~$f'$ of~$G'$ we can extract a proper partial~$q$-coloring~$f$ of~$G[X]$, and if~$f$ cannot be extended to the rest of~$G$ then there is a vertex in~$G - X$ whose neighborhood contains all~$q$ colors; but then such a vertex also exists in~$G'$, contradicting that~$f'$ is proper. 

Observe that the graph~$G'$ consists of the vertex cover~$X$ and the independent set~$G' - X$ of vertices with degree exactly~$q$, no two vertices of which share the same open neighborhood; hence~$G'$ has at most~$k + k^q$ vertices. To prove the lemma we give an efficient encoding for graphs with such a structure. We can represent the instance~$(G', X)$ in~$\Oh(|X|^q) = \Oh(k^q)$ bits as follows. We store an adjacency matrix of~$G'[X]$ in exactly~$k^2$ bits. Then for every set~$S \in \binom{X}{q}$ (of which there are less than~$k^q$) we store exactly one bit, specifying whether or not there is a vertex with open neighborhood~$S$. By fixing some ordering on the vertices of~$X$ for the adjacency matrix, and by fixing an ordering of the sets~$\binom{X}{q}$, the instance can unambiguously be recovered from this encoding. Since the encoding uses less than~$k^2 + k^q$ bits this concludes the proof.
\myqed
\end{proof}

We will investigate richer graph classes which have small \no-certificates for \qlistcoloring in \sectref{section:certificateUpperBounds}, and in \sectref{section:negativeResults} we shall prove that the exponential dependency on the function~$g(q)$ in the kernel size bound is necessary. But first we show that the positive news of \thmref{generalKernel} does not only hold for \qcoloring, but can also be extended to \qlistcoloring under structural parameterizations. For completeness we give a full definition of the generalized form of the problem.

\parproblemdef{$q$-List-Coloring on \fkv graphs}
{An undirected graph~$G$ with for each vertex~$v \in V(G)$ a list~$L(v) \subseteq [q]$ of allowed colors, and a modulator~$X \subseteq V(G)$ such that $G - X \in \f$.}
{The size~$k := |X|$ of the modulator.}
{Is there a proper $q$-coloring $f \colon V(G) \to [q]$ such that $f(v) \in L(v)$ for each~$v \in V(G)$?}

Although the classical complexity of \qcoloring and \qlistcoloring often differs quite significantly - for example, \threecoloring is trivial on bipartite graphs whereas \threelistcoloring is NP-complete on such graphs~\cite[Theorem 1]{BodlaenderJW94} - both coloring problems admit polynomial kernels when parameterized by the vertex-deletion distance to graph classes with small \no-certificates.

\begin{corollary} \label{generalListColoringKernel}
Let \f be a hereditary class of graphs with~$g(q)$-size \no-certificates for \qlistcoloring. The \qlistcoloring problem on \fkv graphs admits a polynomial kernel with $\Oh((k+q)^{q \cdot g(q)})$ vertices for every fixed~$q$.
\end{corollary}
\begin{proof}
Given an instance~$(G,L,X)$ of \qlistcoloring on \fkv graphs, we construct an equivalent instance of \qcoloring on \fkqv graphs as follows. We build~$G'$ by adding a clique on~$q$ vertices~$p_1, \ldots, p_q$ to~$G$, and connect vertex~$p_i$ for~$i \in [q]$ in graph~$G'$ to all vertices~$v$ for which~$i \not \in L(v)$. Since a proper $q$-coloring of~$G'$ assigns unique colors to each of the vertices of the clique, the adjacency of the original vertices to this clique enforces the list requirements. By taking~$X' := X \cup \{p_1, \ldots, p_q\}$ as the new modulator we find that~$G - X = G' - X' \in \f$. Hence the \qcoloring on \fkqv graphs instance~$(G', X')$ is equivalent to the parameterized \qlistcoloring on \fkv graphs instance~$(G,L,X)$. We apply the kernelization algorithm of \thmref{generalKernel} to~$(G',X')$ and since the parameter is~$|X'| = k + q$ the resulting reduced instance satisfies the claimed size bound. Afterwards we assign every vertex a full list~$[q]$ to obtain the kernelized \qlistcoloring instance.
\myqed
\end{proof}

For graph classes which are characterized by a finite set of forbidden induced subgraphs, such as cographs which are induced-$P_4$-free, we can even drop the requirement that a modulator~$X$ is given in the input. A $4$-approximation to the minimum-size modulator to a cograph can be obtained by repeatedly finding an induced~$P_4$ and adding all its vertices to the approximate solution. We may then obtain a polynomial kernel using this approximate modulator as the set~$X$ in \thmref{generalKernel} or \corollaryref{generalListColoringKernel}.

\subsection{Upper bounds to sizes of \no-certificates for \qlistcoloring} \label{section:certificateUpperBounds}
Having proved that an upper bound to the size of \no-certificates for \qlistcoloring on \f yields polynomial kernels for coloring problems on \fkv graphs, we set out to study specific graph classes for which such bounds exist. We start the section by giving an illustrative example for the unions of split graphs. We then consider the more general class containing the union of cochordal graphs, and conclude the section by studying cographs.

\begin{lemma} \label{lemma:splitSmallCertificates}
The class \csplit is hereditary and has $g(q) := q + 4^q$-size \no-certificates for \qlistcoloring.
\end{lemma}
\begin{proof}
Since split graphs are hereditary, so are their unions. To establish the desired bound on the size of \no-certificates, consider a \no-instance~$(G,L)$ of \qlistcoloring with~$G \in \csplit$. If~$G$ contains a clique on~$q + 1$ vertices, then the subinstance induced by this clique cannot be $q$-colored and is therefore a \no-subinstance to list-colorability of size~$q + 1$. In the remainder we may therefore assume~$G$ contains no clique of size more than~$q$. We show how to obtain a \no-subinstance of~$(G,L)$ on at most~$q + 4^q$ vertices. We obtain our \no-subinstance by applying a cleaning rule to~$(G,L)$ which does not affect the list-colorability of the instance.
\begin{cleaningRule} 
If~$(G,L)$ is an instance of \qlistcoloring and there are distinct non-adjacent vertices~$u,v$ such that~$N_G(u) \subseteq N_G(v)$ and~$L(u) \supseteq L(v)$ then~$(G,L)$ is a \yes-instance if and only if $(G - \{u\}, L)$ is a \yes-instance.
\end{cleaningRule}
\begin{proof}
Because~$(G - \{u\}, L)$ is a subinstance of~$(G,L)$ we only have to show that a list-coloring~$f'$ for~$(G - \{u\}, L)$ can be transformed into a list-coloring for~$(G,L)$. Since~$N_G(u) \subseteq N_G(v)$, the color assigned to~$v$ by~$f'$ cannot occur on any neighbor of~$u$, and by the condition~$L(u) \supseteq L(v)$ this color is admissible for~$u$. By non-adjacency of~$u$ and~$v$ it therefore follows that~$f'$ can be extended to a proper list-coloring of~$(G,L)$, by assigning~$u$ the same color as~$v$.
\myqed
\end{proof}
Now we use the cleaning rule to find a small \no-subinstance of our presumed \no-instance~$(G,L)$. Since the rule does not change the answer to the instance, we may exhaustively apply the rule to obtain an equivalent \no-instance~$(G',L')$. We may assume without loss of generality that~$G'$ is connected, because a graph is list-colorable if and only if each connected component is list-colorable; hence in any disconnected \no-instance there is a connected \no-subinstance. We prove that~$|V(G')| \leq q + 4^q$. Because \csplit is hereditary and~$G'$ is connected it follows that~$G' \in \splitgraphs$, so let~$W \dotcup Z \subseteq V(G')$ be a partition of the vertex set of~$G'$ such that~$G'[W]$ is a clique and~$G'[Z]$ is an independent set; this partition exists by the definition of a split graph. By the assumption at the beginning of the proof we know that~$G'$ contains no clique of size more than~$q$, therefore~$|W| \leq q$. Since~$Z$ is an independent set, we know that~$N_{G'}(v) \subseteq W$ for all~$v \in Z$. Consider some~$W' \subseteq W$ and the set of vertices~$S_{W'} := \{ v \in Z \mid N_{G'}(v) = W' \}$. Since all vertices in~$S_{W'}$ are mutually non-adjacent and have the same open neighborhood, by the fact that the cleaning rule is not applicable to~$(G',L')$ it follows that for all vertices~$u,v \in S_{W'}$ we have~$L(u) \not \subseteq L(v)$. In particular this shows that every subset of~$q$ can occur at most once as the list of a vertex in~$S_{W'}$, which proves that~$|S_{W'}| \leq 2^q$. Since all vertices in~$Z$ occur in some set~$S_{W'}$ for~$W' \subseteq W$ it follows that~$|Z| \leq 2^{|W|} 2^q$, and since~$|W| \leq q$ we find~$|Z| \leq 2^{2q}$. As the vertex set of~$G'$ consists of~$W$ and~$Z$ we obtain~$|V(G')| \leq q + 2^{2q} = q+4^q$ which proves the lemma.
\myqed
\end{proof}

Note the unusual use of the cleaning rule in the proof of \lemmaref{lemma:splitSmallCertificates}: although it looks like a kernelization rule for \qlistcoloring, we use it merely to prove the \emph{existence} of small \no-certificates for the class \csplit. The actual kernelization of \qcoloring on \csplitkv graphs implied by \thmref{generalKernel} relies only on the established property, and does not use the cleaning rule.

The next lemma shows that even the class \ccochordal, which is a strict superclass of \csplit, has bounded-size \no-certificates for \qlistcoloring. As the class is more general, the bounding function~$g(q)$ grows faster.

\begin{lemma} \label{lemma:coChordalSmallCertificates}
The class \ccochordal is hereditary and has $g(q) := (q+1)!$-size \no-certificates for \qlistcoloring.
\end{lemma}
\begin{proof}
Since chordal graphs are hereditary, so are their complements and the unions of their complements. As in the previous proof we may restrict our attention to connected graphs without loss of generality, and it therefore suffices to show that for every \no-instance of \qlistcoloring~$(G,L)$ with~$G \in \cochordal$, there is a \no-subinstance with at most~$g(q) := (q+1)!$ vertices. 

We start by investigating the structure of an instance~$(G,L)$ of \qlistcoloring on a cochordal graph. By \proposref{proposition:chordalHasSimplicial} the chordal graph \Gbar has a simplicial vertex~$v$, and from the definition of a simplicial vertex this shows that~$N_{\Gbar}(v)$ is a clique in \Gbar implying that~$N_{\Gbar}(v)$ is an independent set in~$G$. Define the set~$N_{\Gbar}^i(v) := N_{\Gbar}(v) \cap L^{-1}(i)$ containing the vertices which are non-neighbors of~$v$ in~$G$, and which may be assigned color~$i$. Our argument will rely on the following claim.

\begin{claim}
If~$v$ is simplicial in \Gbar and there is a proper $q$-list-coloring of~$(G,L)$ which assigns~$v$ color~$i$, then there is a proper $q$-list-coloring which gives all vertices~$\{v\} \cup N_{\Gbar}^i(v)$ color~$i$.
\end{claim}
\begin{proof}
Consider a proper coloring~$f$ of~$G$ respecting the lists~$L$ with~$f(v) = i$. If there is a vertex~$u \in N_{\Gbar}^i(v)$ with~$f(u) \neq i$, consider the effect of changing~$u$'s color to~$i$. Since~$i \in L(u)$ by definition of~$N_{\Gbar}^i(v)$, we do not violate the requirement that all vertices receive a color from their list. So suppose that changing~$u$'s color to~$i$ results in an improper coloring, because a neighbor of~$u$ was already colored~$i$. Since~$u$ is not a neighbor of~$v$ by definition of~$N_{\Gbar}^i(v) \subseteq N_{\Gbar}(v)$, this means there is a vertex~$w \neq v$ such that $\{u,w\} \in E(G)$ and~$f(w) = i$. Now observe that since~$N_{\Gbar}(v)$ is an independent set in~$G$ and~$u \in N_{\Gbar}^i(v) \subseteq N_{\Gbar}(v)$, the presence of the edge~$\{u,w\}$ shows that~$w \not \in N_{\Gbar}(v)$, i.e., that~$w$ is \emph{not} in the set of~$v$'s non-neighbors. Together with~$w \neq v$ this shows that~$w$ is a neighbor of~$v$ in~$G$. But then~$f$ colors the adjacent vertices~$v$ and~$w$ with color~$i$, which contradicts the assumption that~$f$ is a proper coloring. Hence changing~$u$'s color to~$i$ must result in a proper list coloring, which proves the claim: if~$v$ is colored~$i$, then all vertices in~$N_{\Gbar}^i(v)$ can also receive color~$i$.
\myqed
\end{proof}

Using this claim, we describe an algorithm to solve \qlistcoloring on cochordal graphs. The structure of this algorithm will prove the existence of a small \no-subinstance. The algorithm proceeds as follows on an instance~$(G,L)$. If there is a vertex~$v$ with~$L(v) = \emptyset$, then~$v$ shows that the answer is \no and the algorithm is done. Otherwise we find a vertex~$v$ which is simplicial in~\Gbar, and branch on which color from~$L(v)$ to assign to~$v$. The claim shows that in the branch where we try color~$i \in L(v)$ on~$v$, we may also assign all vertices of~$N_{\Gbar}^i(v)$ color~$i$ without loss of correctness. The answer to an instance in which these vertices receive color~$i$, is equivalent to the answer after removing the vertices~$S_i := \{v\} \cup N_{\Gbar}^i(v)$ from the graph and removing color~$i$ from the lists of the vertices~$N_G(S_i)$. 
Observe that if~$v$ is simplicial in~\Gbar then~$N_G(v) = N_G(\{v\} \cup N_{\Gbar}(v)) = N_G(S_i)$ and therefore we may equivalently simplify the instance in this branch by removing the vertices~$S_i$ and deleting color~$i$ from the lists of the vertices~$N_G(v)$.
The algorithm can therefore safely apply this transformation and call itself recursively on the remaining instance. The crucial insight for the analysis of the algorithm is that in the resulting instance, color~$i$ no longer appears on any list. To see this, note that vertex~$v$ is removed, as are all non-neighbors of~$v$ which had~$i$ on their list. All vertices which are not non-neighbors of~$v$, and are not~$v$ itself, must be neighbors of~$v$; but then color~$i$ was explicitly removed from their lists. Hence in each branch we eliminate at least one color from the instance. Since the instance starts with at most~$q$ different colors occurring in the lists, after branching~$q$ levels deep we must either obtain a graph without vertices (and the answer is \yes), or a graph containing a vertex with an empty list (and the answer is \no).

Using the structure of this algorithm we prove the lemma. Consider a \no-instance~$(G,L)$ of \qlistcoloring where~$G \in \cochordal$. If we run the sketched algorithm on this instance, the answer will be \no since the algorithm is correct. Hence each branch ends with the detection of a vertex with an empty list. Suppose that in each call of the algorithm we mark the simplicial vertex~$v$ that is branched on, and we mark the detected vertex with an empty list when the algorithm outputs \no for a branch. Let~$h(\ell)$ be the number of vertices marked in this process, for an instance whose lists contain~$\ell$ different colors. From this definitions it follows that~$h(0) = 1$. In an instance containing~$\ell$ distinct colors, we mark the simplicial vertex that is branched on and then branch in at most~$\ell$ ways, decreasing the total number of remaining colors in each branch. Hence~$h(\ell) \leq 1 + \ell \cdot h(\ell-1)$. This recurrence is bounded by~$h(\ell) \leq (\ell+1)!$ and since the input instance contains at most~$q$ distinct colors, the total number of vertices marked is at most~$(q+1)!$. 

If we now consider the instance induced by the marked vertices and run the algorithm on this instance, branching on the simplicial vertices in the same order as before, then the algorithm will make exactly the same decisions as for the instance~$(G,L)$; thus the algorithm will output \no, which shows by correctness of the algorithm that the instance induced by the~$(q+1)!$ marked vertices is a \no-subinstance. Since this proves the existence of a \no-subinstance of the requested size, it completes the proof.
\myqed
\end{proof}

By employing a similar argument which shows how the structure of an algorithm to solve \qlistcoloring on cographs can be used to find small \no-subinstances of the problem, we can prove that cographs have bounded-size \no-certificates.

\begin{lemma} \label{lemma:cographsSatisfy}
The class \cograph is hereditary and has $g(q) := 2^{q^2}$-size \no-certificates for \qlistcoloring.
\end{lemma}
\begin{proof}
Since cographs can be characterized as not having an induced~$P_4$ subgraph, it is easy to see they are hereditary. To give a bound on the size of \no-certificates we use a similar approach as in the proof of \lemmaref{lemma:coChordalSmallCertificates}: we give a simple algorithm which correctly decides \qlistcoloring on cographs, and then argue that we can mark at most~$g(q) := 2^{q^2}$ vertices such that the algorithm would also output \no on the instance induced by the marked vertices.

The algorithm for \qlistcoloring uses dynamic programming on a cotree decomposition of the graph (\defref{def:cotree}), similarly to the algorithm of (Klaus) Jansen and Scheffler \cite[Theorem 4.7]{JansenS97}. So let~$(G,L)$ be an instance of \qlistcoloring with~$G \in \cograph$ and consider a cotree representation~$\T$ of the graph~$G$. We create a dynamic programming table~$T[v, S]$ whose first index ranges over the nodes of~$\T$, and whose second index ranges over subsets of~$[q]$. The interpretation of the table is that~$T[v, S] = \true$ if and only if the cograph~$\CG(\T, v)$ represented by the subtree of~$\T$ rooted at~$v$ can be $q$-list-colored with respect to the list assignment~$L$, such that only the colors from~$S$ are used. In this interpretation~$(G,L)$ is a \yes-instance if and only if the root-node~$r$ of~$\T$ satisfies~$T[r, \{1, 2, \ldots, q \} ] = \true$. The table values satisfy the following recurrence:
\begin{equation*}
T[v,S] = 
\begin{dcases}
L(v) \cap S \neq \emptyset & \mbox{If~$v$ is a leaf node.} \\
T[\leftc(v), S] \wedge T[\rightc(v), S] & \mbox{If~$v$ is a \union node.} \\ 
\bigvee _{S' \subseteq S} T[\leftc(v), S'] \wedge T[\rightc(v), S \setminus S'] & \mbox{If~$v$ is a \join node.} \\
\end{dcases}
\label{cographRecurrence}
\end{equation*}
To justify that this recurrence indeed characterizes the behavior of the values~$T[v,S]$, observe that the cograph represented by a leaf~$v$ of a cotree is just a singleton graph on the vertex~$v$ which can be list-colored using colors from~$S$ if~$S$ contains at least one admissible color for~$v$. The graph represented by a cotree whose root is a \union node is simply the disjoint union of the graphs represented by the subtrees rooted at the children. Since no color conflicts can occur between vertices in different connected components of a graph this explains the second item of the recurrence. Finally for the \join node observe that in any proper coloring of a graph~$G_1 \otimes G_2$ there can be no vertex in~$G_1$ which obtains the same color as a vertex of~$G_2$, since they are made adjacent by the \join operation. Hence any proper coloring of~$G_1 \otimes G_2$ using only the color set~$S$ must use some subset~$S' \subseteq S$ for the vertices of~$G_1$ and the disjoint set~$S \setminus S'$ for the vertices of~$G_2$. This proves that a dynamic programming algorithm which computes the values of~$T$ bottom-up in the cotree is correct.

We now use this algorithm to prove that if~$(G,L)$ is a \no-instance of \qlistcoloring with~$G \in \cograph$, then there exists a \no-subinstance on at most~$2^{q^2}$ vertices. Consider a cotree representation~$\T$ of the graph~$G$. If~$G$ contains a clique on~$q+1$ vertices, then this clique cannot be $q$-(list-)colored and hence the subinstance induced by this clique must be a \no-subinstance on~$q+1$ vertices; then we are done. Hence in the remainder we may assume that~$\omega(G) \leq q$, which implies by \proposref{joinHeight} that the join height of~$G$ is smaller than~$q$. Using this fact we now describe a recursive procedure which given a table entry~$T[v,S]$ such that~$T[v,S] = \false$, marks a number of vertices to preserve the \false value of this entry. If~$v$ is a leaf of the cotree, then it suffices to mark that single leaf. If~$v$ is a \union node then at least one of~$T[\leftc(v), S]$ and~$T[\rightc(v), S]$ is \false; recursively mark vertices for a subexpression which yields \false. The most interesting case is a \join node: for all~$S' \subseteq S$, at least one of the terms~$T[\leftc(v), S']$ and~$T[\rightc(v), S \setminus S']$ yields \false. For each subset~$S' \subseteq S$ we recursively mark a ``witness'' for one subexpression which evaluates to \false. 

Let~$M$ be the set of marked vertices and observe that we may obtain a cotree representation of~$G[M]$ by considering the leaves~$L$ of~$\T$ corresponding to the marked vertices, taking the subcotree~$\T'$ induced by the paths from~$L$ to the root, and splicing out the internal vertices of~$\T'$ which have only a single child in~$\T'$. For every node in~$\T'$ there is a corresponding node in~$\T$, and as we have marked witnesses for the \false values in the root of~$\T$, it is not difficult to see that the witnesses ensure \false values in the root of~$\T'$. This shows that the algorithm outputs \no on the subinstance induced by~$M$, and, similarly as in the proof of \lemmaref{lemma:coChordalSmallCertificates}, since the algorithm is correct this proves that the subinstance induced by~$M$ is a \no-instance. It remains to bound the number of marked vertices.

We can bound the number of vertices which were marked to preserve the \no value of a cell~$T[v, S]$ based on the join height of the subtree~$\T_v$ rooted at~$v$. Let~$h(x)$ be the maximum number of vertices which are marked to preserve the \false-value of a cell~$T[v,S]$ with~$\joinheight(\T_v) \leq x$. If~$\joinheight(\T_v) = 0$ then the subtree rooted at~$v$ contains only \union nodes and leaves, and the procedure to mark a vertex for~$T[v,S]$ will trace a path of \union nodes through the tree until it reaches a leaf, and it will mark that single leaf: hence~$h(0) = 1$. Now consider what happens when~$\joinheight(\T_v) > 0$. The marking procedure will trace a path from the root of~$\T_v$, following subexpressions which evaluate to \false, until it reaches a leaf or a \join node. Since we mark only one vertex for a leaf, we focus on the \join node. For this node the procedure will recursively mark vertices for each subset of~$S$. Since~$S \subseteq [q]$ there are at most~$2^q$ subsets, and for each subset we mark vertices in a subtree rooted at~$\leftc(v)$ or~$\rightc(v)$; but note that the join height of these subtrees is at least one lower than that of~$\T_v$. Hence for~$x \geq 1$ we have~$h(x) \leq 2^q \cdot h(x-1)$. We can bound this recurrence as~$h(x) \leq (2^q)^x$. Since we can assume that~$\joinheight(\T) < q$ it follows that the subinstance resulting from marking vertices of~$(G,L)$ contains less than~$h(q) \leq 2^{q^2}$ vertices. This shows the existence of a small \no-subinstance of~$(G,L)$ and concludes the proof.
\myqed
\end{proof}

We note that a more detailed analysis of the marking procedure of the previous theorem gives a better bound. By relating marked vertices to permutations of the members of a partition of the color set, it is possible to prove that the number of marked vertices does not exceed~$q^{3q} = 2^{3q \log q}$. In the interest of brevity we have chosen to give a shorter proof of a weaker bound.

By combining \thmref{generalKernel} and \corollaryref{generalListColoringKernel} with the bounds on sizes of \no-certificates for the hereditary graph classes considered in this section, we obtain the following corollary.

\begin{corollary}
For every fixed integer~$q$, the problems \qcoloring and \qlistcoloring admit polynomial kernels on the parameterized graph classes \csplitkv, \ccochordalkv, and \cographkv.
\end{corollary}

\section{Negative results in the hierarchy} \label{section:negativeResults}
\subsection{Kernelization lower bounds} \label{section:kernelLowerBounds}
We turn to the study of the limits of polynomial kernelizability in our parameter space, to complement the results of the previous section. We first show that \threecoloring on \pathkv graphs does not admit a polynomial kernel unless \containment, thereby presenting a barrier to polynomial kernelizability in the hierarchy of \imgref{ParameterHierarchy}. After establishing this theorem we consider parameterizations for which polynomial kernels were obtained in the previous section, and we give lower bounds on the degree of the polynomial bounding the bitsize of these kernels.

Our first negative result is a superpolynomial kernel lower bound for \threecoloring on \linearforestkv graphs, which we will later lift to the parameterized graph family \pathkv. The gadget in the proof of the following theorem is inspired by a reduction of Lokshtanov et al.~\cite[Theorem 6.1]{LokshtanovMS11}. We learned that Stefan Szeider independently found a similar result for \forestkv graphs.

\begin{theorem} \label{linearForestLowerBound}
\threecoloring on \linearforestkv graphs does not admit a polynomial kernel unless \containment.
\end{theorem}
\begin{proof}
We give a polynomial-parameter transformation (\defref{def:polyParameterTransform}) from \CNFSAT parameterized by the number of variables~$n$ (\proposref{proposition:cnfSatNoPoly}) to \threecoloring parameterized by deletion distance from a linear forest. Consider an input to \CNFSAT which consists of clauses~$C_1, \ldots, C_m$ where each clause is a disjunction of literals of the form~$x_i$ or~$\overline{x_i}$ for~$i \in [n]$. We build a graph~$G$ and a modulator~$X \subseteq V(G)$ such that~$|X| = 2n + 3$ and~$G - X \in \linearforest$.

Construct a clique on three vertices~$p_1, p_2, p_3$; this clique will serve as our palette of three colors, since in any proper $3$-coloring all three colors must occur on this clique. For each variable~$x_i$ for~$i \in [n]$ we make vertices~$T_i$ and~$F_i$ and add the edge~$\{T_i, F_i\}$ to~$G$. We make the vertices~$T_i, F_i$ adjacent to the palette vertex~$p_1$. Now we create gadgets for the clauses of the satisfiability instance.

For each clause~$C_j$ with~$j \in [m]$, let~$n_j$ be the number of literals in~$C_j$ and create a path~$(a^1_j, b^1_j, a^2_j, b^2_j, \ldots, a^{n_j}_j, b^{n_j}_j)$ on~$2n_j$ vertices. We call this the clause-path for~$C_j$. Make the first and last vertices on the path $a_j^1$ and~$b_j^{n_j}$ adjacent to the palette vertex~$p_1$. Make the $b$-vertices $b_j^1, b_j^2, \ldots, b_j^{n_j}$ adjacent to palette vertex~$p_3$. As the last step we connect the vertices on the path to the vertices corresponding to literals. For $r \in [n_j]$ if the~$r$-th literal of~$C_j$ is $x_i$ (resp. $\overline{x_i}$) then make vertex~$a_j^r$ adjacent to~$T_i$ (resp. $F_i$). This concludes the construction of the graph~$G$. We use the modulator~$X := \{T_i, F_i \mid i \in [n]\} \cup \{p_1, p_2, p_3\}$. It is easy to verify that~$|X| = 2n + 3$ and therefore that the parameter of the \threecoloring instance is polynomial in the parameter of \CNFSAT. Since vertices on a clause-path are not adjacent to other clause-paths, it follows that~$G - X$ is a linear forest. It remains to prove that the two instances are equivalent: there is a satisfying assignment for the \CNFSAT instance on clauses~$C_1, \ldots, C_m$ if and only if the graph~$G$ is $3$-colorable.

$(\Rightarrow)$ Assume that~$v \colon [n] \to \{\true,\false\}$ is a satisfying assignment. We construct a proper $3$-coloring~$f \colon V(G) \to [3]$ of~$G$ as follows:
\begin{itemize}
	\item For the palette vertices $p_i$ with $i \in [3]$ define $f(p_i) := i$.
	\item For each~$i \in [n]$ with~$v(i) = \true$, set~$f(T_i) := 2$ and~$f(F_i) := 3$.
	\item For each~$i \in [n]$ with~$v(i) = \false$, set~$f(T_i) := 3$ and~$f(F_i) := 2$.
\end{itemize}
\noindent Using the definition of~$G$ it is easy to verify that this partial coloring~$f$ is proper; it remains to extend~$f$ to the clause-paths. Consider the clause-path~$P_j$ corresponding to a clause~$C_j$. For each vertex~$a_j^r$ whose adjacent literal is \true under~$v$, set~$f(a_j^r) := 3$. Since~$v$ is a satisfying assignment we color at least one vertex on~$P_j$ with~$3$, and since literals which evaluate to \true were given color~$2$ in the previous step, we do not create any conflicts. We now show how to color the remainder of the clause-path~$P_j$. If~$a_j^1$ did not receive a color (i.e.\ its neighboring literal evaluates to \false and the literal-vertex is colored~$3$) then set~$f(a_j^1) := 2$ and~$f(b_j^1) = 1$. Now alternatingly color the successive $a$-vertices with~$2$ and~$b$-vertices with~$1$, until arriving at an $a$-vertex which we already assigned color~$3$ (because its adjacent literal evaluates to \true). This assignment does not create any conflicts. Now start at the last vertex~$b_j^{n_j}$ and color it with~$2$, and work backwards giving uncolored $a$-vertices color~$1$ and~$b$-vertices color~$2$, again until we arrive at a~$3$-colored $a$-vertex. If there are any uncolored subpaths left after this procedure (which occurs if two or more literals of the clause are \true), then color the $a$-vertices on this subpath with~$1$ and the $b$-vertices with~$2$. Using the construction of~$G$ this color assignment is easily seen to be proper. Since the clause-paths are independent, we can perform this procedure independently on each clause-path to obtain a proper~$3$-coloring of~$G$.

$(\Leftarrow)$ Let~$f \colon V(G) \to [3]$ be a proper $3$-coloring of~$G$, and assume without loss of generality (by permuting the color set if needed) that~$f(p_1) = 1, f(p_2) = 2$ and~$f(p_3) = 3$. We show that the \CNFSAT instance has a satisfying assignment. We first show that on every clause-path~$P_j$ corresponding to a clause~$C_j$, there must be a vertex colored~$3$. So assume for a contradiction that some clause-path~$P_j$ is colored using only~$1$ and~$2$. Since the first and last vertices on the path are adjacent to palette vertex~$p_1$, those vertices cannot be colored~$1$ and hence they must be colored~$2$. But since the path has an even number of vertices, \proposref{twoColoringAPath} now shows that the clause-path cannot be $2$-colored using only~$1$ and~$2$ which gives a contradiction. So in a proper $3$-coloring of~$G$ all clause-paths contain a vertex colored~$3$. The $b$-vertices on a clause-path are adjacent to~$p_3$ and hence cannot be colored~$3$; therefore some $a$-vertex~$a_j^r$ which is adjacent to a literal vertex~$T_i$ or~$F_i$ must be colored with~$3$. This implies that the corresponding literal-vertex must be colored~$2$. Now consider the valuation which makes all literals colored~$2$ \true, and all literals colored~$3$ \false. The previous argument shows that at least one literal of each clause is \true. Since~$T_i$ and~$F_i$ are adjacent to each other, and both are adjacent to~$p_1$, we color exactly one of them with~$2$ in a proper~$3$-coloring of~$G$ and hence we obtain a valid satisfying assignment for the \CNFSAT instance.

Since the construction can be carried out in polynomial time and guarantees that the parameter of the output instance is bounded polynomially in the parameter of the input instance, the given reduction is indeed a polynomial-parameter transformation. The theorem now follows from \proposref{proposition:cnfSatNoPoly} since kernel lower bounds are transferred by polynomial-parameter transformations, as noted in \sectref{preliminaries}.
\myqed
\end{proof}

By considering the proof of \thmref{linearForestLowerBound} we can obtain a corollary for a stronger parameterization. Consider the graph~$G$ and modulator~$X$ which is constructed in the proof: the remainder graph~$G - X$ is a linear forest, a disjoint union of paths. By adding vertices of degree two to~$G$ we may connect all the paths in~$G - X$ into a single path. Since degree-2 vertices do not affect the~$3$-colorability of a graph, this does not change the answer to the instance and ensures that~$G - X$ is a single path. Hence we obtain:
\begin{corollary}
\threecoloring on \pathkv graphs does not admit a polynomial kernel unless \containment.
\end{corollary}

The remainder of this section is devoted to the proof of various explicit lower bounds on the degree of the polynomial in the kernel size bounds, using the machinery of Dell and van Melkebeek~\cite{DellM10}. We first consider the \qcoloring problem on \emptykv graphs, giving a lower bound which almost matches the upper bound we obtained earlier, and afterwards we present a general theorem which shows how to construct kernel size lower bounds from large \no-certificates to \qlistcoloring.

Recall that an instance of \qnaesat is a formula built from the conjunction of clauses containing at most~$q$ literals each, which is satisfied if at least one literal in each clause evaluates to \false, and at least one evaluates to \true. By relating \qsat to \qppnaesat (both parameterized by the number of variables) through a folklore reduction we obtain a compression lower bound for \qppnaesat, and by relating the latter to \qppcoloring on \emptykv graphs we can obtain the following theorem.

%
\begin{theorem} \label{coefficientLowerBound}
For every~$q \geq 4$, \qcoloring on \emptykv graphs does not have a kernel of bitsize~$\Oh(k^{q - 1 - \varepsilon})$ for any~$\varepsilon > 0$ unless \containment.
\end{theorem}
\begin{proof}
The proof consists of the following steps. We first show that there is a linear-parameter transformation (\defref{def:polyParameterTransform}) from \qsat parameterized by the number of variables~$n$ to \qppnaesat parameterized by~$n$. As the second step we give a linear-parameter transformation from \qnaesat parameterized by~$n$ to \qcoloring on \emptykv graphs. By combining these two transformations, we can use a kernelization algorithm for \qppcoloring on \emptykv graphs to reduce the size of an instance of \qsat in the following way. Start by transforming a given instance of \qsat on~$n$ variables into an instance of \qppnaesat on~$\Oh(n)$ variables, which in turn is transformed into \qppcoloring on \emptykv graphs with~$k \in \Oh(n)$, and finally apply the kernelization algorithm to this instance. Hence we see that a kernel with~$\Oh(k^{(q + 1) - 1 - \varepsilon})$ bits for \qppcoloring would output an instance of size~$\Oh(n^{q - \varepsilon})$, and by a recent result of Dell and van Melkebeek (\proposref{dellMelkebeekCompressionBound}) such a sparsification algorithm for any~$q \geq 3$ would imply \containment. Hence we find that for~$q \geq 3$ the problem \qppcoloring on \emptykv graphs cannot have kernels with bitsize~$\Oh(k^{(q + 1) - 1 - \varepsilon})$ unless \containment. By a change of variables this shows that for~$q \geq 4$ the \qcoloring on \emptykv graphs problem cannot have kernels with~$\Oh(k^{q - 1 - \varepsilon})$ bits, which implies the theorem. To complete the proof it therefore suffices to give the two linear-parameter transformations.

\begin{claim}
There is a linear-parameter transformation from \qsat parameterized by~$n$ to \qppnaesat parameterized by~$n$.
\end{claim}
\begin{proof}
This transformation can be considered folklore, and was used e.g.\ by Knuth~\cite[Section 6]{Knuth92}; we repeat it here for completeness. Let~$\phi$ be a \qsat-formula on~$n$ variables. Obtain the \qppnaesat formula~$\phi'$ on~$n + 1$ variables by creating a single new variable~$z$, and adding the positive literal~$z$ to each clause of~$\phi$. Any satisfying assignment of~$\phi$ is transformed into a satisfying not-all-equal assignment of~$\phi'$ by setting~$z$ to \false. In the other direction, any not-all-equal assignment which satisfies~$\phi'$ and sets~$z$ to \false, is also a satisfying satisfiability assignment for~$\phi$. But if we have a not-all-equal assignment which sets~$z$ to \true, then we must also obtain a satisfying not-all-equal assignment if we flip the truth assignment of each variable, leading to a not-all-equal assignment which makes~$z$ \false and hence implies that~$\phi$ is satisfiable. Thus the two instances are equivalent, and since~$n' = n + 1$ this constitutes a linear-parameter transformation.
\myqed
\end{proof}
\begin{claim}
There is a linear-parameter transformation from \qnaesat parameterized by~$n$ to \qcoloring on \emptykv graphs.
\end{claim}
\begin{proof}
Consider an instance of \qnaesat on~$n$ variables, and let the clauses be~$C_1, \ldots, C_m$. We build a graph~$G$ and a modulator~$X\subseteq V(G)$ such that~$G-X\in\independent$.

Again we construct a clique to act as our palette, this time using~$q$ colors and corresponding vertices~$p_1,\ldots,p_q$; each vertex will take a different color in each~$q$-coloring of~$G$. 

For each variable~$x_i$ for~$i\in[n]$ we make~$2q$ vertices~$T_{i,1},\ldots T_{i,q},F_{i,1},\ldots,F_{i,q}$. For all~$j \in [q]$ we make~$T_{i,j}$ adjacent to~$F_{i,j}$. Then we make a cycle through successive vertices~$T_{i,1},\ldots,T_{i,q}$ and back to~$T_{i,1}$. We now connect the vertices to the palette, to ensure that there are only two different~$q$-colorings for this gadget (modulo changing the permutation of the colors on the palette). When we discuss the effect of the gadget on potential $q$-colorings for~$G$, we will indicate by color~$i$ ($i\in[q]$) the color that vertex~$p_i$ receives. Now do as follows. For all~$j\in[q]$ we make~$T_{i,j}$ and~$F_{i,j}$ adjacent to all vertices of the palette except vertices~$p_i$ and~$p_{i+1}$, ensuring that $T_{i,j}$ and~$F_{i,j}$ can only take colors~$i$ or~$i+1$. We evaluate these numbers modulo~$q$, e.g.~$T_{i,q}$ is adjacent to all palette vertices except~$p_q$ and~$p_1$.

It is crucial to observe the following about these variable gadgets: each vertex~$T_{i,j}$ can take only two different colors, but each possibility is also one of the only two choices for a neighbor on the cycle. E.g.\  vertex~$T_{i,1}$ can take color~$1$ or~$2$; if it has color~$2$ then vertex~$T_{i,2}$ must take color~$3$, vertex~$T_{i,3}$ must take color~$4$ and so on. Similarly, if~$T_{i,1}$ has color~$1$ then we can make the same argumentation following the cycle in the other direction. Furthermore, since any vertices~$T_{i,j}$ and~$F_{i,j}$ have the same two possible colors~$j$ and~$j+1$ (modulo~$q$), one will take color~$j$ and the other must take color~$j+1$. This means that effectively there are only two possible colorings for the gadget, which are shown in Figure~\ref{figure:colorings}.

\begin{figure}[t]
\centering
\begin{tikzpicture}[thick]
\coordinate (T1) at (1,6);
\coordinate (T2) at (1,5);
\coordinate (T3) at (1,4);
\coordinate (T4) at (1,3);
\coordinate (T5) at (1,2);
\draw (T1)--(T2)--(T3);
\draw[dotted] (T3)--(T4);
\draw (T4)--(T5)..controls (0,2) and (0,6) ..(T1);

\coordinate (F1) at (3,6);
\coordinate (F2) at (3,5);
\coordinate (F3) at (3,4);
\coordinate (F4) at (3,3);
\coordinate (F5) at (3,2);
\draw (T1)--(F1);
\draw (T2)--(F2);
\draw (T3)--(F3);
\draw (T4)--(F4);
\draw (T5)--(F5);

\draw (T1) +(0.5,0.4) node {$T_{i,1}$};
\draw (T2) +(0.5,0.4) node {$T_{i,2}$};
\draw (T3) +(0.5,0.4) node {$T_{i,3}$};
\draw (T4) +(0.5,0.4) node {$T_{i,q-1}$};
\draw (T5) +(0.5,0.4) node {$T_{i,q}$};

\draw (F1) +(0.5,0.4) node {$F_{i,1}$};
\draw (F2) +(0.5,0.4) node {$F_{i,2}$};
\draw (F3) +(0.5,0.4) node {$F_{i,3}$};
\draw (F4) +(0.5,0.4) node {$F_{i,q-1}$};
\draw (F5) +(0.5,0.4) node {$F_{i,q}$};

\draw[fill=white] (T1) circle(0.3cm) node {$1$};
\draw[fill=white] (T2) circle(0.3cm) node {$2$};
\draw[fill=white] (T3) circle(0.3cm) node {$3$};
\draw[fill=white] (T4) circle(0.3cm) node {$q$-$1$};
\draw[fill=white] (T5) circle(0.3cm) node {$q$};

\draw[fill=white] (F1) circle(0.3cm) node {$2$};
\draw[fill=white] (F2) circle(0.3cm) node {$3$};
\draw[fill=white] (F3) circle(0.3cm) node {$4$};
\draw[fill=white] (F4) circle(0.3cm) node {$q$};
\draw[fill=white] (F5) circle(0.3cm) node {$1$};
\end{tikzpicture}
\hspace{2cm}
\begin{tikzpicture}[thick]
\coordinate (T1) at (1,6);
\coordinate (T2) at (1,5);
\coordinate (T3) at (1,4);
\coordinate (T4) at (1,3);
\coordinate (T5) at (1,2);
\draw (T1)--(T2)--(T3);
\draw[dotted] (T3)--(T4);
\draw (T4)--(T5)..controls (0,2) and (0,6) ..(T1);

\coordinate (F1) at (3,6);
\coordinate (F2) at (3,5);
\coordinate (F3) at (3,4);
\coordinate (F4) at (3,3);
\coordinate (F5) at (3,2);
\draw (T1)--(F1);
\draw (T2)--(F2);
\draw (T3)--(F3);
\draw (T4)--(F4);
\draw (T5)--(F5);

\draw (T1) +(0.5,0.4) node {$T_{i,1}$};
\draw (T2) +(0.5,0.4) node {$T_{i,2}$};
\draw (T3) +(0.5,0.4) node {$T_{i,3}$};
\draw (T4) +(0.5,0.4) node {$T_{i,q-1}$};
\draw (T5) +(0.5,0.4) node {$T_{i,q}$};

\draw (F1) +(0.5,0.4) node {$F_{i,1}$};
\draw (F2) +(0.5,0.4) node {$F_{i,2}$};
\draw (F3) +(0.5,0.4) node {$F_{i,3}$};
\draw (F4) +(0.5,0.4) node {$F_{i,q-1}$};
\draw (F5) +(0.5,0.4) node {$F_{i,q}$};

\draw[fill=white] (T1) circle(0.3cm) node {$2$};
\draw[fill=white] (T2) circle(0.3cm) node {$3$};
\draw[fill=white] (T3) circle(0.3cm) node {$4$};
\draw[fill=white] (T4) circle(0.3cm) node {$q$};
\draw[fill=white] (T5) circle(0.3cm) node {$1$};

\draw[fill=white] (F1) circle(0.3cm) node {$1$};
\draw[fill=white] (F2) circle(0.3cm) node {$2$};
\draw[fill=white] (F3) circle(0.3cm) node {$3$};
\draw[fill=white] (F4) circle(0.3cm) node {$q$-$1$};
\draw[fill=white] (F5) circle(0.3cm) node {$q$};
\end{tikzpicture}
\caption{The two colorings for the variable gadget for~$x_i$ corresponding to assigning \true or \false respectively.}\label{figure:colorings}
\end{figure}
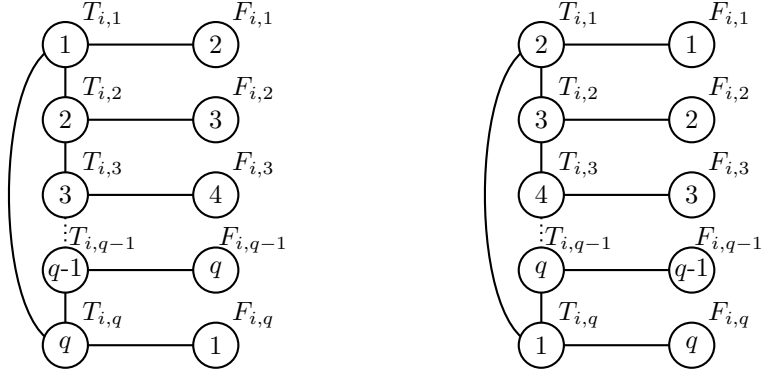

The first coloring, with~$T_{i,1}$ colored~$1$, will be interpreted as assigning \true to~$x_i$, the other one corresponds to assigning \false. We emphasize the key property: if~$x_i$ is \true, then each vertex~$T_{i,j}$ will be colored~$j$ (and each~$F_{i,j}$ will be colored~$j+1$). If~$x_i$ is \false then each~$F_{i,j}$ will be colored~$j$ (and each~$T_{i,j}$ will be colored~$j+1$).

We will now add one vertex~$c_k$ for each clause~$C_k$. Let~$C_k=(\ell_1\vee\ldots\vee\ell_q)$. For each~$\ell_j$ ($j\in[q]$) with~$\ell_j=x_i$, we connect~$c_k$ to~$T_{i,j}$. For each~$\ell_j$ ($j\in[q]$) with~$\ell_j=\neg x_i$, we connect~$c_k$ to~$F_{i,j}$. It is easy to see that this has the desired effect, using the assignment corresponding to a given coloring: if all literals evaluate to \true, then~$c_k$ is connected to one vertex of each color in~$[q]$. Then however,~$c_k$ cannot be properly colored. The same holds when all the literals evaluate to \false.

We briefly sketch correctness of this reduction. If the graph~$G$ can be properly~$q$-colored, then take the assignment corresponding to the coloring of the variable gadgets. The added vertices~$c_k$ ensure that there is no clause with literals all evaluating to \true or all \false.
Conversely, let the formula be satisfiable and color the vertex gadgets according to a satisfying assignment. It follows easily, that among the~$q$ neighbors of any clause vertex~$c_k$ at most~$q-1$ colors are used: indeed, consider two adjacent literals in the clause~$C_k$, i.e.,~$\ell_j$ and~$\ell_{j+1}$, such that~$\ell_j$ evaluates to \false and~$\ell_{j+1}$ to \true. Using Figure~\ref{figure:colorings} it is easy to verify that the neighbor of~$c_k$ that represents~$\ell_j$ is colored~$j+1$ (since~$\ell_j$ evaluates to \false), and the neighbor of~$c_k$ representing~$\ell_{j+1}$ is also colored~$j+1$ (as~$\ell_{j+1}$ evaluates to \true). Hence out of the~$q$ neighbors of~$c_k$ there are two which share the same color, leaving at least one color free for~$c_k$. This implies that the partial coloring can be extended to all clause vertices independently.

Clearly, the clause vertices~$c_k$ form an independent set. Thus we may define the modulator~$X$ as the vertices of the palette together with all vertices of the variable gadgets; the size of this set is~$|X| = 2qn + q$, which is linear in~$n$ for fixed~$q$. It is easy to see that~$G - X \in \independent$. Hence the instance~$(G,X)$ is a valid output of a linear-parameter transformation, which completes the proof.
\myqed
\end{proof}
By the argumentation given at the beginning of the proof, the two claims together prove the theorem.
\myqed
\end{proof}

The proof of \thmref{coefficientLowerBound} shows that an improved compression lower bound for \qnaesat also yields a better lower bound for \qcoloring on \emptykv graphs. In particular, if it would be proven that for~$q \geq 3$ \qnaesat on~$n$ variables cannot be compressed in polynomial time into an equivalent instance on~$\Oh(n^{q - \varepsilon})$ bits, then the kernel of \lemmaref{smallVertexCoverKernel} is optimal up to~$k^{o(1)}$ factors. The remainder of this section revolves around the following notion.

\begin{definition}
Let~\f be a graph class, and let~$(G,L)$ be a \no-instance of \qlistcoloring with~$G \in \f$. Then~$(G,L)$ is an \emph{irreducible \no-instance of \qlistcoloring on \f} if for all vertices~$v \in V(G)$, the subinstance on graph~$G - \{v\}$ is a \yes-instance.
\end{definition}

The following theorem shows how to turn irreducible \no-instances of \qlistcoloring into explicit kernel size lower bounds.

\begin{theorem} \label{theorem:negativeGeneralTheorem}
If \f is closed under disjoint union and there is an irreducible \no-instance of \qminlistcoloring on \f containing~$t \geq 3$ vertices, then \qcoloring on \fkv graphs does not admit a kernel with \emph{bitsize}~$\Oh(k^{t - \varepsilon})$ for any~$\varepsilon > 0$ unless \containment.
\end{theorem}
\begin{proof}
We proceed similarly as in the proof of \thmref{coefficientLowerBound}. We will show how the irreducible \no-instance on~$t$ vertices can be used to give a linear-parameter transformation from \tsat parameterized by the number of variables~$n$ to~\qcoloring on \fkv graphs. So assume that~$(H,L)$ is a \no-instance of \qminlistcoloring with~$|V(H)| = t$ such that any vertex deletion turns it into a \yes-instance. Let the list function~$L$ for this instance assign admissible colors from the set~$[q-2]$, and number the vertices of~$H$ in an arbitrary way as~$h_1, \ldots, h_t$.

Now consider an input to \tsat which consists of clauses~$C_1, \ldots, C_m$, where each clause contains exactly~$t$ literals. Each literal is of the form~$x_i$ or~$\overline{x_i}$ for~$i \in [n]$. We build a graph~$G$ and a modulator~$X \subseteq V(G)$ such that~$|X| = 2n + q$ and~$G - X \in \f$. Construct a clique on~$q$ vertices~$p_1, \ldots, p_q$ to act as a palette. For each variable~$x_i$ for~$i \in [n]$ we make vertices~$T_i$ and~$F_i$ and add the edge~$\{T_i, F_i\}$ to~$G$. Make~$T_i$ and~$F_i$ adjacent to all palette vertices except~$p_{q-1}$ and~$p_q$. We use the irreducible \no-instance~$(H,L)$ to create gadgets for the clauses. For each clause~$C_j$ with~$j \in [m]$, add a disjoint copy of the graph~$H$ to~$G$ and denote it by~$H^j$ on vertices~$h^j_1, \ldots, h^j_t$. For each~$i \in [t]$ make vertex~$h^j_i$ adjacent to all palette vertices~$p_s$ with~$s \in [q-2] \setminus L(h_i)$, and to the vertex~$p_q$. If the $i$-th literal of~$C_j$ is~$x_s$ (resp.~$\overline{x_s}$) then make~$h^j_i$ adjacent to~$T_s$ (resp.~$F_s$). This concludes the description of the graph~$G$. Use the set~$X := \{T_i, F_i \mid i \in [n]\} \cup \{p_1, \ldots, p_q\}$ as the modulator. Since~$G - X$ is a disjoint union of copies of~$H \in \f$, and since~$\f$ is closed under disjoint union by assumption, it follows that~$G - X \in \f$ which proves that~$X$ is indeed a modulator to~$\f$ of size linear in~$n$. Let us now prove the correctness of the reduction: the graph~$G$ is $q$-colorable if and only if the \tsat instance is satisfiable.

($\Rightarrow$) Assume that function~$f \colon V(G) \to [q]$ gives a proper $q$-coloring of~$G$. Since a proper coloring gives unique colors to the vertices of a clique, we may assume without loss of generality that for the palette vertices~$p_i$ with~$i \in [q]$ we have~$f(p_i) = i$. Consider some variable index~$i \in [n]$. By adjacency of the vertices~$T_i$ and~$F_i$ to the palette vertices, and by the edge~$\{T_i, F_i\}$ it follows that one of~$T_i$ is colored~$q-1$ and the other is colored~$q$. Consider the truth assignment which sets a variable~$x_i$ to \true if and only if~$f(T_i) = q$. We will show that this is a satisfying assignment for the input formula.

So consider a clause~$C_j$ of the input formula, which is represented by the graph~$H^j$ which is a copy of~$H$. By adjacency of vertices~$h^j_1, \ldots, h^j_t$ to the palette, it is easy to verify that~$h^j_i$ is assigned a color in~$L(h_i) \cup \{q-1\}$ by the proper coloring~$f$. For the next step, assume for a contradiction that no vertex of the graph~$H^j$ is colored with~$q-1$. Then the coloring induced by~$f$ is a $(q-2)$-list-coloring of~$H^j$ which satisfies the list requirements from~$L$, which contradicts the assumption that~$(H,L)$ is a \no-instance. Hence there is at least one vertex~$h^j_i$ which is colored with~$q-1$. But then the corresponding literal-vertex~$T_s$ or~$F_s$ which is adjacent to~$h^j_i$ cannot be colored~$q-1$, so it is colored~$q$. By our choice of valuation function, the literal evaluates to \true which shows that clause~$C_j$ is satisfied. Since this holds for every clause, the formula is satisfiable.

($\Leftarrow$) Assume that~$v \colon [n] \to \{\true,\false\}$ is a satisfying assignment. We construct a proper $q$-coloring~$f \colon V(G) \to [q]$ of~$G$ as follows:
\begin{itemize}
	\item For the palette vertices $p_i$ with $i \in [q]$ define $f(p_i) := i$.
	\item For each~$i \in [n]$ with~$v(i) = \true$, set~$f(T_i) := q$ and~$f(F_i) := q-1$.
	\item For each~$i \in [n]$ with~$v(i) = \false$, set~$f(T_i) := q-1$ and~$f(F_i) := q$.
\end{itemize}
It is easy to see that this partial assignment is proper; we show how to extend it to the remainder of the graph. The remainder consists of a disjoint copy~$H^j$ for each clause~$j \in [m]$. For each such~$H^j$ function~$v$ satisfies at least one literal of clause~$j$. Suppose that the $i$-th literal is satisfied, and look at the corresponding vertex~$h^j_i$. By the assumption that $(H,L)$ is an irreducible \no-instance of \qminlistcoloring, it follows that the subinstance on graph~$H - \{h_i\}$ is a \yes-instance. Take a $(q-2)$-list-coloring of~$H - \{h_i\}$ and use it for the vertices of~$H^j \setminus \{h^j_i\}$. By the correspondence between the list function and the adjacencies to the palette, and since the literal-vertices~$F_i$ and~$T_i$ only take colors~$q-1$ and~$q$ which are not used in a $(q-2)$-list-coloring, this leads to a proper extension of the coloring~$f$ onto the graph~$H^j \setminus h^j_i$. Observe that at this point, all neighbors of~$h^j_i$ in~$G$ have received a color, and that no such neighbor is colored with~$q-1$: for the neighbors in~$H^j \setminus h^j_i$ this follows from the $(q-2)$-list-coloring, for the palette vertices it follows by construction and for the literal-vertex adjacent to~$h^j_i$ this is ensured by the definition of~$f$. Hence we can safely set~$f(h^j_i) = q-1$ to extend~$f$ onto the entire graph~$H^j_i$. Since the various copies of~$H$ are mutually non-adjacent we can perform this color extension independently for all copies to obtain a proper coloring of~$G$.

This concludes the correctness proof of the linear-parameter transformation. It is easy to see that it can be computed in polynomial time. To establish the theorem, observe that if \qcoloring on \fkv graphs would have a kernel with bitsize~$\Oh(k^{t - \varepsilon})$, then we could compress an instance of \tsat on~$n$ vertices by transforming it to the \qcoloring on \fkv graphs instance with parameter value~$k \in \Oh(n)$, and then applying the kernelization algorithm to obtain an equivalent instance of \qcoloring of bitsize~$\Oh(n^{t - \varepsilon})$. Since \qcoloring is obviously decidable, by \proposref{dellMelkebeekCompressionBound} a kernel of such bitsize does not exist unless \containment.
\end{proof}

As a graph on~$\Oh(k^{t/2 - \varepsilon})$ vertices can be encoded in~$\Oh(k^{t/2 - \varepsilon})^2 = \Oh(k^{t - 2 \varepsilon})$ bits using its adjacency matrix, \thmref{theorem:negativeGeneralTheorem} gives the following corollary.
\begin{corollary} \label{corollary:lowerBoundsFromIrreducibleInstances}
If \f is closed under disjoint union and there is an irreducible \no-instance of \qminlistcoloring on \f containing~$t \geq 3$ vertices, then \qcoloring on \fkv graphs does not admit a kernel with~$\Oh(k^{t/2 - \varepsilon})$ \emph{vertices} for any~$\varepsilon > 0$ unless \containment.
\end{corollary}
In the next section we will prove the existence of large irreducible \no-instances to \qlistcoloring to supply explicit kernel size lower bounds.

\subsection{Lower bounds to the sizes of \no-certificates for \qlistcoloring} \label{section:certificateLowerBounds}
In \sectref{section:certificateUpperBounds} we constructed functions which bound the sizes of \no-certifi\-cates for \qlistcoloring on the graph classes \csplit, \ccochordal, and \cograph. The bounds we obtained were exponential in the number of colors~$q$ used in the instances. In this section we prove that the exponential dependence on~$q$ cannot be avoided, by giving explicit constructions of large irreducible \no-instances. We start by showing that for the class of paths, there is \emph{no} function depending on~$q$ alone which can bound the size of \no-certificates, which explains why there is no polynomial kernel for the \threecoloring problem on \pathkv graphs.

\begin{lemma}
For every~$q \geq 2$ and every even integer~$t \geq 2$ there is an irreducible \no-instance of \qlistcoloring on a path graph, containing~$t$ vertices.
\end{lemma}
\begin{proof}
Let~$G$ be the path on successive vertices~$v_1, \ldots, v_t$ with~$t \geq 2$ an even integer. Define~$L(v_1) = L(v_t) := \{1\}$, and let~$L(v_i) := \{1,2\}$ for~$1 < i < t$. We claim that~$(G,L)$ is an irreducible \no-instance of \qlistcoloring. To see that it is a \no-instance, observe that~$v_1$ has only one color on its list and must therefore be colored~$1$, which forces~$v_2$ to be colored~$2$, which forces~$v_3$ to be colored~$1$, and so on: coloring~$v_1$ with~$1$ forces all even-numbered vertices to be colored~$2$. But this prevents~$v_t$ from being assigned a color on its list, since~$t$ is even and~$L(v_t) = \{1\}$. After a single vertex~$v_i$ is removed from the path, we may find a proper $q$-list-coloring by alternatingly assigning~$1$ and~$2$ from the beginning of the path until reaching~$v_i$, and also alternating~$1$ and~$2$ backwards starting from~$v_t$. Hence~$(G,L)$ is indeed an irreducible \no-instance.
\myqed
\end{proof}

There is one unified construction which simultaneously acts as a lower bound for the three considered graph classes \csplit, \ccochordal, and \cograph.

\begin{lemma} \label{lemma:certificateLowerBound}
For every even integer~$q \geq 2$ there is an irreducible \no-instance of \qlistcoloring on a graph in the set $\splitgraphs \cap \cochordal \cap \cograph$, containing~$q / 2 + \binom{q}{q/2}$ vertices.
\end{lemma}
\begin{proof}
Let~$q$ be even and consider the \qlistcoloring instance constructed as follows. Initialize~$G$ as a clique~$X$ on~$q/2$ vertices and set~$L(x) := [q]$ for all~$x \in X$. For every~$S \in \binom{[q]}{q/2}$ add a vertex~$v_S$ to~$G$ with~$N_G(v_S) := X$ and~$L(v_S) := S$. Let the resulting instance be~$(G,L)$, and observe that~$G$ is a split graph on~$q/2 + \binom{q}{q/2}$ vertices since it partitions into an independent set and a clique. Since split graphs are chordal and closed under complementation we have~$G \in \cochordal$. To see that~$G$ is a cograph observe that it is the join of a clique and an independent set, both of which are cographs, and that cographs are closed under taking joins. We will prove that~$(G,L)$ is irreducible: it is a \no-instance to \qlistcoloring, but any vertex removal turns it into a \yes-instance.

We first show that~$(G,L)$ is a \no-instance. Assume for a contradiction that~$f \colon V(G) \to [q]$ is a proper list-coloring of~$G$. Since~$X$ is a clique, all vertices of~$X$ receive different colors under~$f$. Let~$S := f^{-1}(X)$ be the~$|X| = q/2$ colors assigned to~$X$ by~$f$. During the construction of~$G$ we added a vertex~$v_S$ with neighborhood~$X$ and list~$L(v_S) := S$ to the graph. But the choice of~$v_S$ ensures that all colors of~$L(v_S)$ occur on the neighbors of~$v_S$, and therefore~$f$ cannot be a proper list-coloring; contradiction. Hence~$(G,L)$ is indeed a \no-instance.

Now we prove that any vertex-deletion turns~$(G,L)$ into a \yes-instance. So let~$z \in V(G)$ be an arbitrary vertex, and consider the subinstance~$(G - \{z\},L')$ where~$L'$ is the restriction of~$L$ to the remaining vertices. 
\begin{itemize}
	\item If~$z \in X$, then assign the~$q/2 - 1$ remaining vertices of the clique~$X' := X - \{z\}$ colors~$1$ up to~$q/2 - 1$; this is compatible with their lists. Now consider the remaining vertices. For each~$S \in \binom{[q]}{q/2}$ the vertex~$v_S$ has a list of~$|S| = q/2$ admissible colors, and after removal of~$z$ vertex~$v_S$ has only~$q/2 - 1$ neighbors left in~$G - \{z\}$. Hence there must be one color on the list of~$v_S$ which is not yet used on a neighbor, and we use that color for~$v_S$. Since the vertices not in~$X$ form an independent set, we may assign these colors independently to obtain a proper $q$-list-coloring of~$(G - \{z\},L')$. 
	\item Now consider the remaining case that~$z \not \in X$, so~$z = v_S$ for some~$S \in \binom{[q]}{q/2}$. Give each vertex of~$X$ a unique color from the set~$S$. For each set~$S' \in \binom{[q]}{q/2} \setminus \{S\}$ it remains to choose a color for the vertex~$v_{S'}$. By definition we have~$L'(v_{S'}) = S'$. The neighbors of~$v_{S'}$ in~$G - \{v_S\}$ are exactly the vertices of the clique~$X$ and they use up all colors from~$S$; so the colors in~$S' \setminus S$ can be used on~$v_{S'}$. Since~$S' \neq S$ and~$|S'| = |S| = q/2$ it follows that there is at least one color in~$S' \setminus S$ which is available for use on~$v_{S'}$, and as before we can use the fact that the vertices not in~$X$ form an independent set to argue that this leads to a proper $q$-list-coloring of the instance~$(G - \{z\}, L')$.
\end{itemize}
Since this shows that~$(G,L)$ is an irreducible \no-instance with~$G \in \splitgraphs \cap \cochordal \cap \cograph$ this concludes the proof.
\end{proof}

An overview of the various upper and lower bounds for \no-certificate and kernel sizes can be found in \tableref{tab:overview}. We conclude the section with the following corollary, which results from combining the previous lemma with \corollaryref{corollary:lowerBoundsFromIrreducibleInstances}.

\begin{corollary}
For every even integer~$q \geq 4$ and~$\varepsilon > 0$ the \qcoloring problem on parameterized \csplitkv, \ccochordalkv, or \cographkv graphs does not admit a kernel with~$\Oh \left(k^{\frac{1}{2}(q/2 - 1 + \binom{q-2}{q/2-1}) - \varepsilon} \right)$ vertices unless \containment.
\end{corollary}

\section{Domination-related parameters} \label{section:dominationRelated}
In this section we show that the complexity of \threecoloring is strongly related to the domination-structure of the graph.

\begin{theorem} \label{threeColoringByDomSet}
\threecoloring on a general graph~$G$ can be solved in~$\Oh^*(3^k)$ time when given a dominating set~$X$ of size~$k$.
\end{theorem}
\begin{proof}
Let~$X$ be a dominating set in graph~$G$ of size~$k$. The algorithm proceeds as follows. For each of the~$3^k$ possible assignments of colors to~$X$, we check in linear time whether adjacent vertices received the same color. If the partial coloring is proper then we determine whether it can be extended to the remainder of the graph, and this check can be modeled as a \threelistcoloring instance on the graph~$G - X$: for every vertex~$v \in G - X$ the list of available colors is formed by those elements of~$\{1,2,3\}$ which do not occur on~$v$'s neighbors in~$X$. Since~$X$ is a dominating set, every vertex has at least one colored neighbor and therefore each vertex of~$G - X$ has a list of at most two available colors. It has long been known that such \threelistcoloring instances can be solved in polynomial time by guessing a color for a vertex and propagating the implications; see for example the survey by Tuza~\cite[Section 4.3]{Tuza97}. Hence for each assignment of colors to~$X$ we can test in polynomial time whether it can be extended to~$G - X$ or not, and~$G$ is $3$-colorable if and only if at least one of these attempts succeeds.
\myqed
\end{proof}

On graph classes such as planar graphs and apex-minor-free graphs~\cite{FominLRS11} where \domset can be suitably well approximated, the requirement that a dominating set is given in the input can be dropped from \thmref{threeColoringByDomSet} at the cost of an increase in the running time, showing that \threecoloring is fixed-parameter tractable on these graph classes when parameterized by the domination \emph{number}. Whether such approximation is possible on general graphs, even in FPT-time, is one of the important open problems of parameterized approximation theory~\cite{Marx08}.

The fixed-parameter tractability of \threecoloring parameterized by the size of a given dominating set raises the question whether the problem admits a polynomial kernel. Assuming \ncontainment this is not the case, which can be seen from the proof of \thmref{linearForestLowerBound}: the modulator~$X$ which is constructed in the proof is a dominating set in the graph, and therefore the given reduction serves as a polynomial-parameter transformation from \CNFSAT parameterized by~$n$ to \threecoloring parameterized by the size of the dominating set~$X$.

Having a restricted domination-structure in a \threecoloring instance \emph{does} make it more amenable to kernelization, which becomes clear when we use a different parameter. Recall that the class \dominated contains those graphs in which each connected component has a dominating vertex.
\begin{theorem} \label{threeColoringDominated}
\threecoloring on \dominatedkv graphs admits a polynomial kernel.
\end{theorem}
\begin{proof}
The outline of the proof is as follows. We first give a reduction procedure which transforms an instance of \threecoloring on \dominatedkv graphs to a graph in an even simpler graph class. Let \simpledominated be the class of graphs in which every connected component~$C$ has a dominating vertex~$v$ such that~$C - \{v\}$ has maximum degree at most one. We show that an instance~$(G, X)$ on \dominatedkv graphs can be reduced to an instance~$(G',X')$ on \simpledominatedkv graphs. As the second step we show that the class \simpledominated satisfies all requirements of \thmref{generalKernel}. Therefore an instance of \threecoloring on \dominatedkv graphs can be kernelized by transforming it to \simpledominatedkv, and then applying \thmref{generalKernel}. Since~$\simpledominated \subseteq \dominated$ the output instance of on \simpledominatedkv graphs is also a valid, equivalent instance on \dominatedkv graphs which shows that this suggested kernelization indeed outputs an instance of the \emph{same} problem as was given in the input. To prove the theorem it therefore suffices to supply the two mentioned ingredients.
\begin{lemma}
There is a polynomial-time algorithm which transforms an instance of \threecoloring on \dominatedkv graphs into an equivalent instance on \simpledominatedkv graphs.
\end{lemma}
\begin{proof}
We sketch the actions of the algorithm. On input~$(G, X)$ the algorithm first tests whether~$G$ contains an odd wheel as a subgraph - recall that an odd wheel is the graph obtained by adding a universal vertex to a cycle on an odd number of vertices. The test can be done in polynomial time by noting that~$G$ contains an odd wheel if and only if there is a vertex~$v \in V(G)$ such that~$G[N(v)]$ is not bipartite. It is easy to see that at least four colors are required to properly color an odd wheel; hence if we find an odd wheel in~$G$ then the answer to the instance is \no and we output a constant-sized \no-instance.

As the next step, the algorithm tests whether~$G - X$ contains a diamond as a subgraph. Recall that a diamond is a graph on vertices~$\{u,x,y,v\}$ with edges~$\{ \{u,x\}, \{u,y\}, \{v, x\}, \{v,y\}, \{x,y\} \}$. Observe that if~$G - X$ contains a diamond subgraph on~$\{u,x,y,v\}$ then the edge~$\{u,v\}$ cannot be present, otherwise this would be an odd wheel with a rim of three vertices (i.e.~$K_4$), and it would have been found by the previous stage. In any proper $3$-coloring of the diamond the vertices~$u$ and~$v$ must receive the same color: if they obtain different colors, then the two colors used on~$u$ and~$v$ cannot be used on~$x$ or~$y$ (by their adjacencies), but~$x$ and~$y$ must also receive different colors from each other since~$\{x,y\} \in E(G)$; this cannot be done with only three colors. From the fact that~$u$ and~$v$ are non-adjacent and must receive the same color in any~$3$-coloring, it is not hard to see that~$G$ is $3$-colorable if and only if the graph~$G'$ which is obtained from~$G$ by identifying~$u$ and~$v$ is $3$-colorable. Hence we may transform the instance~$(G,X)$ to~$(G', X)$. Observe that after this identification we have~$G' - X \in \dominated$: this follows easily from the definition of \dominated, using the fact that we have identified non-adjacent vertices~$u,v$ from the \emph{same} connected component of the graph~$G - X \in \dominated$. After the modification to the graph the algorithm starts over by testing for an odd wheel. The algorithm halts when no diamond is found in~$G - X$; since each reduction step reduces the number of vertices there are at most~$|V(G)|$ rounds until termination and the whole process takes polynomial time. 

To conclude the proof we show that the output~$(G, X)$  of the reduction algorithm must satisfy~$G - X \in \simpledominated$. Since the input graph minus its modulator is contained in~$\dominated$ and each reduction step preserves this fact, the output satisfies~$G - X \in \dominated$. Additionally, the graph~$G - X$ is diamond-subgraph-free when no more reduction steps apply to it, and does not contain an odd wheel. To prove that such a graph~$G - X$ is contained in \simpledominated, consider a connected component~$C$ of~$G - X$. Since~$G - X \in \dominated$ there is a dominating vertex~$u$ in~$C$. 
If~$C - \{u\}$ contains a vertex~$v$ of degree more than two (i.e.\ not counting the neighbor~$u$), then we derive a contradiction. So let~$v \in N_G(u)$ have at least two more neighbors~$x,y \in N_G(v) \setminus \{u\}$; since~$u$ is a dominating vertex in~$C$ the vertices~$x,y$ are also adjacent to~$u$.
If~$\{x,y\} \in E(G)$ then the vertices~$\{u,x,y,v\}$ form a~$K_4$ and hence an odd wheel; if~$\{x,y\} \not \in E(G)$ then~$\{u,x,y,v\}$ is a diamond. Since the output of the reduction algorithm is diamond-subgraph-free and odd-wheel-subgraph-free, every connected component~$C$ of~$G - X$ must have a dominating vertex~$u$ such that~$C - \{u\}$ has degree at most one: thus~$G - X \in \simpledominated$ which concludes the proof.
\myqed
\end{proof}
\begin{lemma}
\simpledominated graphs satisfy the conditions of \thmref{generalKernel} and therefore \qcoloring on \simpledominatedkv graphs admits a polynomial kernel for every fixed~$q$.
\end{lemma}
\begin{proof}
To see that the class \simpledominated is hereditary, consider a graph~$G \in \simpledominated$ and the effect of deleting a vertex~$v$ from a connected component~$C$ in the graph~$G$. If~$v$ is an isolated vertex then it is easy to see from the definition that~$G - \{v\} \in \simpledominated$. Otherwise, let~$u$ be a dominating vertex in component~$C$ such that~$C - \{u\}$ has maximum degree at most one, which exists since~$G \in \simpledominated$. If~$u \neq v$ then the vertex~$u$ remains a dominating vertex for~$C$, and the maximum degree of~$C \setminus \{u\}$ does not increase by deletion of~$v$. If~$u = v$ then by definition~$G[N(v)]$ has maximum degree at most one, which implies that~$C - \{v\}$ is a disjoint collection of paths on one or two vertices. Since each such path satisfies the conditions of a graph in \simpledominated this shows that \simpledominated is hereditary.

It remains to prove that for any \no-instance~$(G,L)$ of \qlistcoloring on a graph~$G \in \simpledominated$ there is a \no-subinstance on at most~$f(q) := 2q + 1$ vertices. We show how to mark at most~$2q + 1$ vertices such that the answer to the subinstance induced by the marked vertices is \no. Assume without loss of generality that~$G$ is connected, and let~$u$ be a dominating vertex such that~$G - \{u\}$ has maximum degree at most one; such a vertex exists by the definition of \simpledominated. Since~$G - \{u\}$ has maximum degree at most one, it is a collection of paths on at most two vertices. 

For each possible color assignment to~$u$, it is not possible to properly color the remaining vertices using colors from their lists since~$(G,L)$ is a \no-instance. Hence there is at least one component of~$G - \{u\}$ to which the coloring cannot extended, given the color of~$u$. Mark the (at most two) vertices in one such component, and also mark the vertex~$u$. We mark at most two vertices for each of the~$q$ possible color assignments to~$u$, and we mark~$u$; hence we mark at most~$2q + 1$ vertices. It is easy to see that the marked vertices induce a \no-subinstance, which concludes the proof.
\myqed
\end{proof}
Having supplied the two necessary ingredients, the proof of \thmref{threeColoringDominated} is completed.
\myqed
\end{proof}

\section{Conclusions}


\begin{table}
	\centering
	\small
		\begin{tabular}{@{}lllll@{}}
	\toprule
\multirow{2}{2cm}{Parameterized graph family} & \multicolumn{2}{c}{\no-certificate bound} & \multicolumn{2}{c}{Kernel bitsize bound} \\ \cmidrule(r){2-3} \cmidrule(l){4-5}
& upper & lower & upper & lower \\
\midrule
\emptykv & 1 & 1 & $\Oh(k^q)$ & $\Omega(k^{q - 1 - \varepsilon})$ \\
\pathkv & $\infty$ & $\infty$ & $\Oh(q^k)$ & No~$\Oh(k^{f(q)})$ \\
\csplitkv & $q + 4^q$ & $q/2 + \binom{q}{q/2}$ & $\Oh(k^{2q(q+4^q)})$ &  $\Omega(k^{q/2-1+\binom{q-2}{q/2-1}-\varepsilon})$ \\
\ccochordalkv & $(q+1)!$ & $q/2 + \binom{q}{q/2}$ & $\Oh(k^{2q (q+1)!})$ &  $\Omega(k^{q/2-1+\binom{q-2}{q/2-1}-\varepsilon})$ \\
\cographkv & $2^{q^2}$ & $q/2 + \binom{q}{q/2}$ & $\Oh \left(k^{2 q \cdot 2^{q^2}} \right)$ & $\Omega(k^{q/2-1+\binom{q-2}{q/2-1}-\varepsilon})$ \\
\bottomrule
		\end{tabular}
	\caption{Kernel bounds for \qcoloring on parameterized graph families. The table shows upper- and lower bounds for the sizes of \no-certificates for \qlistcoloring on graphs in \f, and upper- and lower bounds for the number of bits in kernels for \qcoloring on \fkv graphs. In the order of the listed graph families, the upper bound on the number of bits needed for kernel representations follows a) from \lemmaref{smallVertexCoverKernel}, b) from the existence of an~$\Oh^*(q^k)$-time FPT algorithm~\cite{BodlaenderK08}, and c,d,e) by squaring the number of vertices that results from \thmref{generalKernel}. The kernel size upper bounds holds for all values of~$q$, whereas the lower bounds hold for all even~$q \geq 4$; they are obtained using the size of a \no-certificate for \qminlistcoloring with \thmref{theorem:negativeGeneralTheorem}.}
	\label{tab:overview}
\end{table}

We studied the kernelizability of \qcoloring within a hierarchy of structural parameterizations of the input graph, obtaining several positive and negative results. Our positive results extend to \qlistcoloring, and of course the negative results do as well since \qlistcoloring is strictly more general than \qcoloring.

Ever since the first paper on parameterized coloring problems by Cai, it has been known that the complexity of \listcoloring on a graph class \f and the parameterized complexity of \chromaticnumber on \fkv graphs are strongly related: if \listcoloring on \f is polynomial-time solvable then the resulting parameterized problem lies in FPT~\cite[Theorem 3.3]{Cai03a}. Marx gave a refined view of the interaction between \precoloringextension on \f under parameterizations relating to the number of precolored vertices, and the complexity of \chromaticnumber under structural parameterizations. The results in this paper give another example of such a connection. We have shown that for graph classes \f closed under disjoint union and vertex deletion, the existence of polynomial kernels for \qcoloring on \fkv graphs is governed by the behavior of \no-certificates for \qlistcoloring on graphs in \f: when there is an upper bound~$g(q)$ on the size of \no-subinstances then there is a kernel with~$\Oh(k^{q \cdot g(q)})$ vertices, whereas a lower bound on the size of \no-subinstances to \qminlistcoloring in the form of an explicit irreducible \no-instance on~$t$ vertices, shows that no kernel exists with $\Oh(k^{t/2 - \varepsilon})$ vertices for any~$\varepsilon > 0$ unless \containment.

To apply our general kernelization upper and lower bound theorems, we proved bounds on the sizes of \no-certificates for \qlistcoloring on the graph classes \pathgraphs, \csplit, \ccochordal, and \cograph. For the class of paths, no function of~$q$ can bound the size of \no-certificates. For other classes we were able to establish the existence of such functions, which grow exponentially with~$q$. The lower bounds show that this exponential dependence on~$q$ is necessary, and that therefore the numbers of vertices in kernels for the considered parameterized coloring problems must grow very rapidly with the number of colors used: the degree of the polynomial which represents the kernel size must be exponential in~$q$. It would be interesting to see whether the results we obtained regarding the sizes of \no-certificates for \qlistcoloring have applications in the field of certifying algorithms (cf.~\cite{BruceHS09}). \tableref{tab:overview} gives an overview of the bounds which were obtained for the various parameterized graph classes.

We conclude with some directions for further work. One could try to find larger classes of graphs with bounded-size \no-certificates for \qlistcoloring. Since such a class cannot contain all paths, a natural candidate would be the class of~$P_5$-free graphs, a superclass of cographs in which \qcoloring is known to be polynomial-time solvable for every fixed~$q$~\cite{HoangKLSS10}. Sandwiched between the cographs and the~$P_5$-free graphs, one might also consider~$P_4$-sparse graphs or related graph classes~\cite[Section 11.4]{BrandstadtLS99}.

The parameter hierarchy we considered uses vertex-deletion distance to well-studied graph classes as the parameter. The question of kernelizability can also be asked for the edge-deletion and edge-completion variants of these parameters~\cite{Cai03a,Marx06a} which will result in quite a different boundary between tractable and intractable: it is not hard to see that \qcoloring on \linearforestpmke graphs admits a polynomial kernel by deleting vertices of degree at most two, whereas \thmref{linearForestLowerBound} shows that this is not the case on \linearforestkv graphs. As a final open question it will be interesting to settle the gap between the kernelization upper and lower bounds of \qcoloring on \emptykv graphs.

\textbf{Acknowledgments.}
We are grateful to an anonymous referee of STACS 2011~\cite{BodlaenderJK11} who suggested investigating Dell and van Melkebeek-type lower bounds for \qcoloring parameterized by vertex cover.

\bibliography{FullInfo}
\bibliographystyle{elsarticle-num}

\end{document}